\documentclass[a4paper,11pt]{extarticle}
\pdfoutput=1 

\usepackage{jheppub} 

\usepackage[T1]{fontenc} 
\usepackage{float}

\usepackage{xcolor}

\newcommand{\hz}{\mathit{1}}
\newcommand{\ho}{\mathit{2}}
\newcommand{\htw}{\mathit{3}}
\newcommand{\hth}{\mathit{4}}

\newcommand{\ev}[1]{\langle #1 \rangle}

\title{\boldmath Describing metric-affine theories anew: alternative frameworks, examples and solutions}

\author[a,b]{Damianos Iosifidis}
\author[a,1]{Konstantinos Pallikaris,\note{Corresponding author.}}

\affiliation[a]{Laboratory of Theoretical Physics, Institute of Physics, University of Tartu, W. Ostwaldi 1, 50411
Tartu, Estonia}
\affiliation[b]{Institute of Theoretical Physics, Department of Physics Aristotle University of Thessaloniki, 54124 Thessaloniki, Greece}

\emailAdd{damianos.iosifidis@ut.ee}
\emailAdd{konstantinos.pallikaris@ut.ee}

\abstract{In this work we describe metric-affine theories anew by making a change of field variables. A series of equivalent frameworks is presented and identifications are worked out in detail. The advantage of applying the new frameworks is that any MAG theory can be handled as a Riemannian theory with additional fields. We study the Hilbert-Palatini action using the new field variables and disclose interesting symmetries under $SO$ transformations in field space. Then, we use solvable and suitable Riemannian theories as seed models for solvable MAG theories, restricting ourselves to three examples. We present a black hole solution with torsion and non-metricity which under a certain tuning acquires a regular core. A de Sitter universe with the expansion powered by 3-form torsion, is also reported.}

\begin{document} 
\maketitle
\flushbottom

\section{\label{sec:Intro}Introduction}

The general theory of relativity is perhaps one of the most elegantly simple theories of physics with such a strong impact. The ingenious statement that matter tells space-time how to curve, and curved space-time tells matter how to move, crystallized into Einstein's equations, \[ G_{\mu\nu}=\frac{8\pi G}{c^4}T_{\mu\nu}, \] along with the many experimental tests the theory has successfully passed since its birth, have truly established it as the most widely accepted theory of gravitation.

Despite its successes, Einstein's theory has its shortcomings. To mention but a few of them, first, general relativity is not a perturbatively renormalizable quantum field theory meaning that it gets striped of its predictive power at high energies with the Planck mass constituting the cut-off scale. Second, the small measured value of the cosmological constant leads to a perplexing discrepancy between theory and experiment, a naturalness problem known as the cosmological constant problem (see~\cite{Martin:2012bt} for an extensive review). Prominent shortcomings are also the flatness and horizon problems~\cite{PhysRevLett.22.1071,dicke1970gravitation,1979grec.conf..504D,Weinberg:1972kfs} plaguing the standard model of hot big bang cosmology, which are properly addressed in cosmic inflation~\cite{PhysRevD.23.347}.

In view of the above, looking for alternative gravity theories is a justified course of action.\footnote{For a review of the zoo of modified-gravity theories see~\cite{CANTATA:2021ktz,clifton2012modified,Heisenberg:2018vsk} and references therein.} The search for these so-called modified theories of gravity is in essence a search for healthy field equations that differ from those of Einstein. Owing to Lovelock and his undisputed theorem~\cite{Lovelock:1971yv,1972JMP....13..874L}, there is a list of assumptions that we need to break (in one or more ways) in order to find such a set of equations. In particular, one of the assumptions is that space-time is a smooth manifold equipped with a Lorentzian metric and a unique affine connection, induced by the latter, known as the Levi-Civita connection. Therefore, one of the many ways to dodge the stringent consequences of the theorem, is to permit the affine connection to be an independent field variable beyond the metric.

A general connection has both torsion and non-metricity, and a gravitational theory for the metric and the affine connection is known as a {Metric-Affine Gravity} (MAG) theory~\cite{HEHL19951}. MAG theories exhibit many attractive features. First, the presence of a new gravitational potential, the independent affine connection, brings gravity conceptually closer to the other interactions whose mediators are gauge connections.\footnote{See the notion of affine gauge theory in~\cite{HEHL19951}.} Second, an intriguing feature of metric-affine theories of gravity is the emergence of a hypermomentum current~\cite{HehlKerlickHeyde+1976+111+114,HehlKerlickHeyde+1976+524+527,HehlKerlickHeyde+1976+823+827,HEHL1976446} in the presence of matter couplings to the gauge connection. 

This differential form, obtained by varying the matter action with respect to the gauge connection, can be decomposed into the irreducible spin, dilation and shear parts which ought to excite the post-Riemannian structure. In the above sense, MAG theories bring forth an astonishing interplay between matter with non-trivial microstructure and non-Riemannian effects. Finally, note that an interesting discussion has been revived about the status of MAG as a quantum theory though definitive conclusions are yet far from being drawn (see \cite{Percacci:2020ddy,Percacci:2020bzf,Pagani:2015ema} and references therein).\footnote{See also~\cite{Iosifidis:2019dua,Iosifidis:2020gth,Iosifidis:2021nra,Shimada:2018lnm,Mikura:2020qhc,BeltranJimenez:2020sqf,PhysRevD.101.044011,Iosifidis:2018zwo,Iosifidis:2021fnq,Iosifidis:2021bad,Rigouzzo:2022yan,Jimenez-Cano:2022sds,Bahamonde:2021akc,BeltranJimenez:2020sih,Kubota:2020ehu} for some recent advances in the field.}

As with Riemannian theories, one is particularly interested in MAG theories whose field equations can be solved, ideally providing exact solutions (at least for some symmetry ansatz). If the task of finding exact solutions in Riemannian theories is in most cases a difficult one, then the trouble gets double in MAG because we also have to determine the connection. In fact, perhaps the most persistent obstruction to obtaining exact solutions with non-vanishing torsion and non-metricity in metric-affine theories,\footnote{See~\cite{Bahamonde:2022kwg,Bahamonde:2020fnq,Tresguerres:1995js,Tucker:1995fw,Obukhov:1996pf} and references therein for some examples of black hole solutions with torsion and/or non-metricity.} is the computational complexity one is bound to face when attempting to solve the field equations for the affine connection. 

In the dominant part of the MAG literature, the strategy to make the connection dynamical is, roughly put, to consider (at least) quadratic curvature invariants like $R^2$, $R_{\mu\nu}R^{\mu\nu}$, $R^{\lambda\rho\mu\nu}R_{\lambda\rho\mu\nu}$, et cetera. This strategy is indeed well-motivated and fairly general, but it can quickly turn any attempt at finding a solution into a nearly impossible task, even for relatively simple (in form) Lagrangians of this sort. The reason behind this is that the affine connection is a very compact package of a large number of degrees of freedom, the dynamics of which are encoded in the components of a single tensor equation and presented in an awfully coupled manner. In fact, among other techniques, one almost always tries to split this master tensor equation into simpler, hopefully decoupled equations by acting on it with some symmetry projector, or taking traces. Therefore, it may not always be the case that the affine connection is the optimal field variable, beyond the metric, to describe a MAG theory, at least not for all intents and purposes.   

In this work, we embrace this point of view and use it as a motivation for our proposal. Our goal is to make a change of field variables that will allow us to trade the connection field equations for an equivalent ``decongested'' system of simpler field equations obtained by letting an action vary with respect to tensor and vector fields. These tensor and vector fields, used to describe MAG theories anew, will be the irreducible pieces of torsion and non-metricity under the Lorentz group. Ipso facto, they are identified with the fundamental fields, the metric and the affine connection. We then work out a complete mapping between the two frameworks which can be later used as a dictionary. The advantage of the new framework is that we can now handle MAG theories as Riemannian theories with additional fields, at least within the context of the variational problem. These additional fields are part of the space-time geometry it self, and not some external entities. 

With the new framework established, we proceed with giving examples of how to construct MAG theories in vacuum which result in a selective and tractable self-excitation of the connection. Although there is no universal prescription, a basic idea underlies all our examples. We take Riemannian theories with additional fields (vectors and tensors), which we exactly know how to solve, and we cast them, after some minor necessary modifications, into MAG theories which \emph{effectively} yield the same field equations. The role of the additional fields is now performed by the new field variables. Their propagation is tantamount to the excitation of (part of) the post-Riemannian structure. Even though the form of the metric solution in such MAG theories will, more or less, be already known in the gravity literature, the full solution, including the connection, will be novel, for it will in general feature non-zero torsion/non-metricity backgrounds.   

\paragraph{Plan of this work.} In section~\ref{sec:Prelim} we convey the bare minimum in metric-affine theories. Then, in section~\ref{sec:AF} we present the alternative framework and a detailed mapping between the latter and the ordinary Palatini approach. Using the new framework, we revisit the Hilbert-Palatini action in section~\ref{sec:HilbertNew} hoping for fresh insight, and we introduce a useful variant of the new framework when projective symmetry is at play. Finally, in section~\ref{sec:Excited} we showcase a series of examples where we apply the previously developed frameworks, and we also report solutions therein, concluding in section~\ref{sec:Conclusions}.

\section{\label{sec:Prelim}Preliminaries}
This section is devoted to a brief communication of the MAG preliminaries. In metric-affine theories the affine connection is an independent field variable beyond the metric. We use it to define a covariant derivative whose action on vector and co-vector fields is given by 
\begin{subequations}
\begin{eqnarray}
  \nabla_\mu V^\nu &= \partial_\mu V^\nu + \Gamma^\nu{}_{\lambda\mu}V^\lambda,\\
  \nabla_\mu V_\nu &= \partial_\mu V_\nu- \Gamma^\lambda{}_{\nu\mu}V_\lambda,
\end{eqnarray}
\end{subequations}
where $\Gamma^\lambda{}_{\mu\nu}$ are the connection symbols. A general affine connection features both torsion and non-metricity given by
\begin{subequations}\label{def:Tor+Nm}
\begin{eqnarray}
  T^\lambda{}_{\mu\nu}&=&2\Gamma^\lambda{}_{[\nu\mu]},\\
  Q_{\lambda\mu\nu}&=&-\nabla_\lambda g_{\mu\nu},
\end{eqnarray}
\end{subequations}
respectively. The former introduces twisting; parallel transport along a closed path results in a translation. The latter measures the failure of the metric to be covariantly constant; parallel transport brings about a change in vector norms. 

Out of torsion and non-metricity we can construct three vectors and one axial tensor. Regarding torsion, we have the vector $T_\mu = T^\lambda{}_{\mu\lambda}$ and the axial tensor
\begin{equation}
S^{\alpha_1...\alpha_{n-3}} = -\frac{1}{6(n-3)!}\tilde{\epsilon}^{\alpha_1...\alpha_{n-3} \lambda \mu \nu} T_{\lambda \mu \nu}.\label{def:AxialTensor}
\end{equation}
Here, $\tilde{\epsilon}_{\alpha_1 ... \alpha_n}=\sqrt{-\mathsf g}\epsilon_{\alpha_1 ... \alpha_n}$ with $\epsilon_{\alpha_1 ... \alpha_n}$ being the Levi-Civita symbol in $n$ space-time dimensions. Our convention for the symbol is $\epsilon_{01...n-1}=1=-\epsilon^{01...n-1}$. In $n=4$ dimensions, the above axial tensor is known as the torsion pseudo-vector, 
\begin{equation}
\label{def:PseudoVector}
S^{\alpha} = -\frac{1}{6}\tilde{\epsilon}^{\alpha \lambda \mu \nu} T_{\lambda \mu \nu}.
\end{equation}
Regarding non-metricity, we have the vector $Q_{\mu}=Q_{\mu\alpha\beta}g^{\alpha\beta}$, which is proportional to what is often called the Weyl vector in MAG lore, and $
\check{Q}_{\mu}=Q_{\alpha\beta\mu}g^{\alpha\beta}$.

Continuing, we define the curvature tensor of the general affine connection as
\begin{equation}\label{def:Curvature}
{R^\mu}_{\nu \alpha \beta} =  \partial_{\alpha} \Gamma^\mu{}_{\nu\beta} +  \Gamma^\mu{}_{\rho\alpha} \Gamma^\rho{}_{\nu\beta}- \alpha\leftrightarrow \beta.
\end{equation}
From the above we can form three independent contractions, 
\begin{subequations}
    \begin{eqnarray}
        R_{\nu \beta} &=& {R^\mu}_{\nu \mu \beta},\\
\hat{R}_{\alpha \beta} &=& {R^\mu}_{\mu \alpha \beta} = \partial_{[\alpha} Q_{\beta]},\\
\check{R}^\lambda{}_{\alpha} &=& {R^\lambda}_{\mu \alpha \nu} g^{\mu \nu},
    \end{eqnarray}
\end{subequations}
which go by the name Ricci tensor, homothetic-curvature tensor, and co-Ricci tensor, respectively. Notice that only the last contraction requires a metric. Finally, contracting indices once more with the metric, we form the Ricci scalar  $
R= R_{\mu \nu} g^{\mu \nu} =  \check{R}^{\mu}{}_\mu$. As per tradition, we will refer to the curvature tensor associated with the Levi-Civita connection as the Riemann tensor. Its single (double) trace will bear the name Riemannian Ricci tensor (scalar).

Furthermore, it is a well-established fact that every affine connection differs from another affine connection by a tensor. Therefore, we can always write a general affine connection as
\begin{equation}\label{eq:AffineDecomposition}
\Gamma^\lambda{}_{\mu \nu} = \tilde{\Gamma}^\lambda{}_{\mu \nu} + {N^\lambda}_{\mu \nu},
\end{equation}
where
\begin{equation}\label{def:LeviCivita}
\tilde{\Gamma}^\lambda{}_{\mu\nu} = \frac12 g^{\rho\lambda}\left(\partial_{\mu} 
g_{\nu\rho}+\partial_{\nu} 
g_{\mu\rho} - \partial_\rho g_{\mu\nu}\right)
\end{equation}
are the Christoffel symbols, and 
\begin{eqnarray}\label{eq:DistortionExp}
{N^\lambda}_{\mu\nu} &=& \frac{1}{2} g^{\rho\lambda}\left(Q_{\mu\nu\rho} + Q_{\nu\rho\mu}
- Q_{\rho\mu\nu} -T_{\rho\mu\nu} -
T_{\nu\mu\rho} - T_{\mu\nu\rho}\right)
\end{eqnarray}
is the so-called distortion tensor encompassing the non-Riemannian DoF. Torsion and non-metricity can always be traded for the distortion tensor via the relations $
 T^\lambda{}_{\mu\nu} = -2N^\lambda{}_{[\mu\nu]}$ and $
Q_{\lambda\mu\nu}= 2 N_{(\mu \nu )\lambda}$.

Note that eq.~\eqref{eq:AffineDecomposition} suggests that we can split off any quantity into a Riemannian part and non-Riemannian contributions; this is the reputed post-Riemannian expansion of a quantity. For instance, the post-Riemannian expansion of the curvature tensor reads
\begin{equation}
R^\mu{}_{\nu \alpha \beta}  = \tilde{R}^\mu{}_{ \nu \alpha \beta} + 2 \tilde{\nabla}_{[\alpha} {N^\mu}_{|\nu|\beta]} + 2 {N^\mu}_{\lambda[\alpha} {N^\lambda}_{|\nu|\beta]},
\end{equation}
where $\tilde{\nabla}_{\alpha}$ is the Levi-Civita covariant derivative and $\tilde{R}^\mu{}_{ \nu \alpha \beta} $ the Riemann tensor. Unless otherwise stated, quantities with a tilde accent will always stand for objects associated with the Levi-Civita connection.

\section{\label{sec:AF}The alternative framework}

Observe that the presence of an affine connection as an independent field variable introduces $n^3$-many additional \emph{a priori} DoF. Undeniably, the affine connection, being an essential constituent of the metric-affine geometry, is a meaningful variable to work with; torsion and non-metricity are after all properties of a connection. However, squashing that many degrees into a single field is not always the most convenient option. In this section, we instead distribute them among seven fields which correspond to the irreducible pieces of torsion and non-metricity.\footnote{A similar logic was adopted in~\cite{Rigouzzo:2022yan} where the authors split torsion and non-metricity into tensor and trace parts.} This strange way of re-organizing the connection DoF will be suitable for purposes presented during a later stage. Let us also remark that the use of these field variables is essentially in accordance with the distortion tensor becoming the key variable, instead of the connection. 

The new fields will of course be identified with the metric and the affine connection, the fundamental field variables in metric-affine theories, thus allowing us --- via this change of field variables --- to describe any MAG theory anew. We will show in full generality that the field equations derived within this new framework imply and are implied by the field equations obtained in the familiar context of the Fundamental (or Palatini) Framework (FF) where the metric and the affine connection are the independent variables. The freedom to switch between different formulations of the same theory will prove to be a great asset in the next sections. 

In what follows, $\mathring{a}{}_{\lambda\mu...}$ denotes the completely traceless part of a tensor $a_{\lambda\mu...}$,  whereas $\Bar{a}{}_{\lambda\mu...}$ denotes the complement of $\mathring{a}{}_{\lambda\mu...}$ in $a_{\lambda\mu...}$, viz., $\Bar{a}{}_{\lambda\mu...}=a_{\lambda\mu...}-\mathring{a}{}_{\lambda\mu...}$. The irreducible decomposition of the torsion tensor under the Lorentz group yields 
\begin{equation}\label{eq:TorsionIRD}
    T_{\lambda\mu\nu}=H_{\lambda\mu\nu}+\mathring{t}{}_{\lambda\mu\nu}+\bar{t}{}_{\lambda\mu\nu},
\end{equation}
where 
    \begin{eqnarray}\label{eq:TorsionIRDconstituents}
        H_{\lambda\mu\nu}&=&T_{[\lambda\mu\nu]},\qquad {t}{}_{\lambda\mu\nu}=T_{\lambda\mu\nu}-H_{\lambda\mu\nu}.
    \end{eqnarray}
In particular, 
\begin{equation}
        \bar{t}{}_{\lambda\mu\nu}=\frac{2}{n-1}g_{\lambda[\nu}T_{\mu]}.
\end{equation}
Note that instead of the 3-form field $H_{\lambda\mu\nu}$ one may alternatively use the dual tensor $S^{\alpha_1 ... \alpha_{n-3}}$ defined in~\eqref{def:AxialTensor}. 

Similarly, for non-metricity we have
\begin{equation}\label{eq:NmIRD}
    Q_{\lambda\mu\nu}=\mathring{\pi}{}_{\lambda\mu\nu}+\bar{\pi}{}_{\lambda\mu\nu}+\mathring{q}{}_{\lambda\mu\nu}+\bar{q}{}_{\lambda\mu\nu},
\end{equation}
where 
\begin{equation}\label{eq:NmIRDconstituents}
    \pi_{\lambda\mu\nu} = Q_{(\lambda\mu\nu)},\qquad q_{\lambda\mu\nu}=Q_{\lambda\mu\nu}-\pi_{\lambda\mu\nu}.
\end{equation}
In particular,
\begin{equation}
    \bar{\pi}{}_{\lambda\mu\nu}=\frac{1}{n+2}g_{(\lambda\mu}\rho_{\nu)},\qquad \bar{q}_{\lambda\mu\nu}=\frac{2}{3(n-1)}\left( g_{\lambda(\mu}u_{\nu)}-g_{\mu\nu}u_\lambda\right),
\end{equation}
with
\begin{equation}
    u_\mu = \Check{Q}_\mu - Q_\mu,\qquad
        \rho_\mu = 2 \Check{Q}_\mu + Q_\mu.
\end{equation}

Using the defining eqs.~\eqref{def:Tor+Nm}, equations~\eqref{eq:TorsionIRDconstituents} and~\eqref{eq:NmIRDconstituents} tell us how to express the irreducible pieces in terms of the metric and the affine connection. The other way around, eqs.~\eqref{eq:TorsionIRD} and~\eqref{eq:NmIRD} tell us how to express the affine connection in terms of the metric and the irreducible pieces using eqs.~\eqref{eq:AffineDecomposition},~\eqref{def:LeviCivita}, and~\eqref{eq:DistortionExp}.  

Since we will work with many fields, we find it befitting to use multi-field notation. Let us introduce two objects, $O$ and $A$, with components $O_{\lambda\mu\nu}^{N}$ and $A_\mu^{I}$, respectively. They are given by 
\begin{subequations}\label{def:multi-fields}
\begin{eqnarray}
    {O}_{\lambda\mu\nu}&=&\left\lbrace H_{\lambda\mu\nu} , {t}_{\lambda\mu\nu} , {\pi}_{\lambda\mu\nu} , {q}_{\lambda\mu\nu}\right\rbrace,\\
    {A}_\mu &=& \left\lbrace T_\mu , \rho_\mu , u_\mu\right\rbrace.
\end{eqnarray}
\end{subequations}
Einstein's summation convention will also be adopted for indices $M,N,...$, which take values in $\{\hz,\ho,\htw,\hth\}$, and for indices $I,J,...$, which take values in the subset $\{\ho,\htw,\hth\}$.\footnote{Note the use of slanted numerals for the value of an internal index as opposed to $\mu,\nu,... = 0,..n-1$.} We can lower/raise these indices with the reference metrics $\delta_{MN}$ and $\delta_{IJ}$, respectively. As above, whenever the capital indices are omitted, the objects should be understood as column vectors in Euclidean space. Finally, the term Alternative Framework (AF) will be coined for the formulation of a MAG theory in terms of the set $\{ g,\mathring{O}^N,A^I\}$ of field variables.

With all the necessary ingredients at our disposal, let us consider a general $n$-dimensional MAG action in the FF, say
  \begin{equation}
  I[g,\Gamma] =\int \sqrt{-\mathsf g}d^nx\mathcal{L},\label{eq:GeneralAction}
\end{equation}
where $\mathsf{g}\equiv \det g$. We let it vary in order to get 
\begin{equation}\label{eq:FFvariation}
  \delta I=\int\sqrt{-\mathsf g}d^nx\left(E_{\mu\nu}\delta g^{\mu\nu} +\Delta_\lambda{}^{\mu\nu}\delta\Gamma^\lambda{}_{\mu\nu}\right)+ \text{s.t.},
\end{equation}
where s.t. denotes the surface terms arising from integrating by parts. We have also abbreviated the functional derivatives as
\begin{equation}
    {E}_{\mu\nu}=\frac{1}{\sqrt{-\mathsf g}}\frac{\delta I}{\delta g^{\mu\nu}}, \qquad {\Delta}_\lambda{}^{\mu\nu}=\frac{1}{\sqrt{-\mathsf g}}\frac{\delta{I}}{\delta\Gamma^\lambda{}_{\mu\nu}}
\end{equation}
The field equations read 
\begin{equation}
  E_{\mu\nu}=0,\qquad \Delta_\lambda{}^{\mu\nu}=0,\label{eq:FFFE}
\end{equation}
with ${E}_{\mu\nu}$ being a symmetric tensor.\footnote{\label{foot:Surface}The delicate issue of surface-term handling is out of the scope of this paper. We rather assume that one has by all means ensured that the variational problem is well-posed.}

On the other hand, considering eqs.~\eqref{eq:AffineDecomposition},~\eqref{def:LeviCivita},~\eqref{eq:TorsionIRD}, and~\eqref{eq:NmIRD}, we can write the previous action in the AF, namely
\begin{equation}
    I\left[g,\mathring{O}{}^{N},A^{I}\right]=\int\sqrt{-\mathsf{g}}d^nx\mathcal{L}.
\end{equation}
Letting $I$ vary we get 
\begin{equation}\label{eq:AFvariation}
    \delta I = \int \sqrt{-\mathsf{g}}d^nx \left( \hat{E}_{\mu\nu}\delta g^{\mu\nu} +\mathring{\mathcal{O}}{}^{\lambda\mu\nu}_N \delta \mathring{O}{}_{\lambda\mu\nu}^{N} + \mathcal{A}^\mu_{I}\delta A_\mu^{I} \right) 
\end{equation}
plus surface terms where $\mathring{\mathcal{O}}{}_{N}$ and $\mathring{O}{}_{N}$ belong to the same irreducible tensor subspace as Lorentz tensors. The field equations read 
\begin{equation}\label{eq:AFFE}
    \hat{E}_{\mu\nu}=0,\qquad \mathring{\mathcal{O}}^{\lambda\mu\nu}_{N}= 0,\qquad \mathcal{A}^\mu_{I}= 0,
\end{equation}
where $\hat{E}_{\mu\nu}$ is a symmetric tensor. Observe that the traceless property of $\mathring{O}{}_{\lambda\mu\nu}^{N}$ must be preserved when the action is varied. This condition can be enforced with a Lagrange multiplier. The result is equivalent to simply demanding that the functional derivative
with respect to $\mathring{O}{}_{\lambda\mu\nu}^{N}$, $\mathring{\mathcal{O}}^{\lambda\mu\nu}_{N}$, must be traceless.

With the above in hand, we turn our attention to finding the identities relating the functional derivatives $\{\hat{E}, \mathring{\mathcal{O}}{}^{N},\mathcal{A}^{I}\}$ to $\{E,\Delta\}$. These identities will arise via identifications. Expressing eq.~\eqref{eq:FFvariation} in terms of the AF variables, one finds that
\begin{subequations}\label{eq:AFtoFFFEcon}
    \begin{eqnarray}
        \mathring{\mathcal{O}}{}^{\lambda\mu\nu}_{\hz}&=&-\frac12 \Delta^{[\lambda\mu\nu]},\qquad
        \mathring{\mathcal{O}}{}^{\lambda\mu\nu}_{\ho}=\mathring{D}{}^{[\mu\nu]\lambda},\\
        \mathring{\mathcal O}{}^{\lambda\mu\nu}_{\htw}&=&\frac12 \mathring{\Delta}{}^{(\lambda\mu\nu)},\qquad
        \mathring{\mathcal O}{}^{\lambda\mu\nu}_{\hth}=-\mathring{D}{}^{\lambda(\mu\nu)},\\
        {\mathcal{A}}^\mu_{\ho}&=&\frac{1}{n-1}\left( \Delta^{\mu\lambda}{}_\lambda-\Delta_\lambda{}^{\mu\lambda}\right),\\
        \mathcal{A}^\mu_{\htw} &=& \frac{1}{6(n+2)}\left(\Delta_\lambda{}^{\lambda\mu}+\Delta_\lambda{}^{\mu\lambda}+\Delta^{\mu\lambda}{}_{\lambda}\right),\\
        \mathcal{A}^\mu_{\hth} &=& \frac{1}{3(n-1)}\left( 2\Delta^{\mu\lambda}{}_\lambda-\Delta_\lambda{}^{\lambda\mu}-\Delta_\lambda{}^{\mu\lambda}\right),
    \end{eqnarray}
\end{subequations}
where $\mathring{D}$ and $\mathring{\Delta}$ are given in eqs.~\eqref{eq:DeltaIRDconstituents} of the appendix. Finally, we also have
\begin{eqnarray}\label{eq:AFtoFFFEmet}
    \hat{E}_{\mu\nu}&=& E_{\mu\nu} -\frac{\Delta^{\alpha }{}_{(\mu \nu )} -  \delta^{\alpha }_{(\mu }\Delta_{\nu )\beta }{}^{\beta }}{n-1}\left[\frac{n-1}{6(n+2)}\rho_\alpha+T_\alpha+\frac{2}{3}u_\alpha\right]+\frac{2\Delta^{[\alpha\beta]}{}_\beta }{n-1}\mathring{t}{}_{(\mu\nu)\alpha}-\nonumber\\
    &&-\frac{ 2\Delta^{(\alpha\beta)}{}_\beta + \Delta^{\beta}{}_{\beta}{}^\alpha}{2(n+2)}\mathring{\pi}{}_{\mu\nu\alpha}+\frac{  \Delta^{\alpha\beta}{}_\beta-\Delta_\beta{}^{(\alpha\beta)}}{n-1}\mathring{q}{}_{\alpha\mu\nu}-\nonumber\\
    &&- \frac12 {\Delta_{(\mu}{}^{\alpha\beta}\left( H_{\nu) \alpha \beta }- \mathring{\pi}{}_{\nu) \alpha \beta }+2\mathring{q}{}_{\nu)\alpha\beta}+2\mathring{t}{}_{|\beta\alpha|\nu)}\right)}+\nonumber\\
    &&+\frac{1}{2} \left(\tilde{\nabla}^{\alpha }\Delta_{(\mu \nu )\alpha } + \tilde{\nabla}^{\alpha }\Delta_{(\mu |\alpha |\nu )} -  \tilde{\nabla}^{\alpha }\Delta_{\alpha (\mu \nu )}\right),
\end{eqnarray}
where we used the identities 
\begin{equation}
    \mathring{t}{}_{[\mu\nu]\lambda}=-\frac12 \mathring{t}{}_{\lambda\mu\nu},\qquad \mathring{q}{}_{(\mu\nu)\lambda}=-\frac12 \mathring{q}{}_{\lambda\mu\nu}.
\end{equation}
If we let the fields $\mathring{O}^N$ and $A^I$ on the right hand side of eq.~\eqref{eq:AFtoFFFEmet} denote expressions involving the metric and the connection symbols (see eqs.~\eqref{eq:TorsionIRDconstituents} and~\eqref{eq:NmIRDconstituents}), the above simply gives us $\hat{E}$ in terms of the FF quantities. 

At this stage, we find it useful to display the ``inverted form'' of eqs.~\eqref{eq:AFtoFFFEcon} by expressing $\Delta$ in terms of $\mathring{\mathcal{O}}{}_{N}$ and $\mathcal{A}_{I}$. Using the identity
\begin{equation}
    \mathring{D}{}_{[\mu\nu]\lambda}=\mathring{D}{}_{[\mu|\lambda|\nu]}-\mathring{D}{}_{\lambda[\mu\nu]},
\end{equation}
we directly obtain
\begin{subequations}\label{eq:FFtoAFFEpre1}
    \begin{eqnarray}
        \mathring{\Delta}{}^{[\lambda\mu\nu]} &=& -2\mathring{\mathcal O}{}^{\lambda\mu\nu}_{\hz},\qquad
        \mathring{\Delta}{}^{(\lambda\mu\nu)} = 2\mathring{\mathcal O}{}^{\lambda\mu\nu}_{\htw},\\
        \mathring{D}{}^{\lambda[\mu\nu]} &=& -2\left( \mathring{\mathcal O}{}^{[\mu\nu]\lambda}_{\hth}+\mathring{\mathcal O}{}^{\lambda\mu\nu}_{\ho}\right),\qquad
        \mathring{D}{}^{\lambda(\mu\nu)} = -\mathring{\mathcal O}{}^{\lambda\mu\nu}_{\hth}.
    \end{eqnarray}
\end{subequations}
The last three equations in~\eqref{eq:AFtoFFFEcon} form a separate matrix subsystem, invertible for $n>1$, whose inversion yields
\begin{subequations}\label{eq:FFtoAFFEpre2}
    \begin{eqnarray}
        \Delta_\lambda{}^{\lambda\mu}&=&(n-1)\left( \mathcal{A}^\mu_{\ho}  - 2 \mathcal{A}^\mu_{\hth}\right)+ 2(n+2) \mathcal{A}^\mu_{\htw},\\
        \Delta_\lambda{}^{\mu\lambda}&=&(n-1)\left( \mathcal{A}^\mu_{\hth}  - \mathcal{A}^\mu_{\ho}\right)+ {2(n+2)} \mathcal{A}^\mu_{\htw},\\
        \Delta^{\mu\lambda}{}_\lambda&=&(n-1)\mathcal{A}^\mu_{\hth} + 2(n+2)\mathcal{A}^\mu_{\htw}.
    \end{eqnarray}
\end{subequations}

Recalling that the irreducible decomposition of a general rank-3 tensor has the form~\eqref{eq:DeltaIRD}, it all boils down to the equation
\begin{eqnarray}\label{eq:FFtoAFFEcon}
    \Delta^{\lambda\mu\nu}/2&=&\mathring{\mathcal O}{}_{\htw}^{\lambda\mu\nu}-\mathring{\mathcal O}{}_{\hz}^{\lambda\mu\nu}+\mathring{\mathcal O}{}^{\nu\mu\lambda}_{\hth}-\mathring{\mathcal O}{}_{\ho}^{\lambda\mu\nu}+  \nonumber\\
    &&+3\mathcal{A}^{(\lambda}_{\htw}g^{\mu\nu)}+g^{\nu(\lambda}\mathcal{A}^{\mu)}_{\hth} - g^{\mu\lambda}\mathcal{A}^\nu_{\hth}+g^{\lambda[\mu}\mathcal{A}^{\nu]}_{\ho}.
\end{eqnarray}
Finally, one may take the above result, plug it into eq.~\eqref{eq:AFtoFFFEmet}, and write the latter as 
\begin{equation}\label{eq:FFtoAFFEmet}
    E_{\mu\nu}=\hat{E}_{\mu\nu}+...,
\end{equation}
which provides us with $E$ in terms of AF quantities. Finally note that the vanishing of a tensor implies that all its irreducible pieces vanish separately, and vice versa.

Let us now mold all these technical details into the main result we wish to convey. When the field eqs.~\eqref{eq:AFFE} hold true, we have that $\Delta_\lambda{}^{\mu\nu}=0$ via eq.~\eqref{eq:FFtoAFFEcon} and $E_{\mu\nu}=0$ via eq.~\eqref{eq:FFtoAFFEmet}, ergo, the field eqs.~\eqref{eq:FFFE} are implied. The other way around, when the field eqs.~\eqref{eq:FFFE} hold true, we have $\mathring{\mathcal O}{}^{\lambda\mu\nu}_{N}=0$ and $\mathcal{A}^\mu_{I} = {0}$ via eqs.~\eqref{eq:AFtoFFFEcon}. From eq.~\eqref{eq:AFtoFFFEmet}, it further follows that $\hat{E}_{\mu\nu}=0$, ergo, the field eqs.~\eqref{eq:AFFE} are implied. Consequently, we have shown an equivalence relation in detail, in particular, that the field equations in the two formulations {imply and are implied by} each other. We also remark that, having the field equations in one of the two frameworks, it is always possible to reconstruct the field equations in the other. 

Lastly, having set up the new framework, we find it useful to report an interesting correspondence. There exist certain linear connection transformations in the FF which amount to translations of only one irreducible piece at a time (preserving the rest) in the AF. Before disclosing them, let us bring yet another pair of multi-fields to our aid, $o$ and $a$, with components ${o}_{\lambda\mu\nu}^{N}$ and $a_\mu^{I}$, respectively. Note that $\mathring{o}{}^{N}$ and $\mathring{O}^{N}$ belong to the same irreducible tensor subspace as Lorentz tensors. After some straightforward algebra we arrive at a 1:1 correspondence between the translations
    \begin{eqnarray}\label{eq:ShiftsAF}
      \mathring{O}{}'^{N} = \mathring{O}{}^{N} + \mathring{o}{}^{N},\qquad A'{}^{I}=A^I + a^I,
    \end{eqnarray}
in the AF (space-time indices understood, thus omitted) and the linear connection transformations $\Gamma'^{\lambda}_{\mu\nu}=\Gamma^\lambda_{\mu\nu}+(\delta\Gamma)^\lambda_{\mu\nu}$ with
  \begin{subequations}\label{eq:LCTs}
    \begin{eqnarray}
      (\delta\Gamma)_{\lambda\mu\nu}&=&-\frac{1}{2}\mathring{o}{}_{\lambda\mu\nu}^{\hz},\qquad
        (\delta\Gamma)_{\lambda\mu\nu}=-\mathring{o}{}_{\nu\mu\lambda}^{\ho},\\
     (\delta\Gamma)_{\lambda\mu\nu}&=&\frac{1}{2}\mathring{o}{}_{\lambda\mu\nu}^{\htw},\qquad
      (\delta\Gamma)_{\lambda\mu\nu}=-\mathring{o}{}_{\lambda\mu\nu}^{\hth},\\
    (\delta\Gamma)_{\lambda\mu\nu}&=&\frac{2}{n-1}g_{\nu[\mu}a_{\lambda]}^{\ho},\qquad
       (\delta\Gamma)_{\lambda\mu\nu}=\frac{1}{2(n+2)}a^{\htw}_{(\lambda}g_{\mu\nu)},\\
      (\delta\Gamma)_{\lambda\mu\nu}&=&\frac{2}{3(n-1)}\left(g_{\mu\nu}a^{\hth}_{\lambda}-g_{\lambda(\mu} a_{\nu)}^{\hth}\right),
    \end{eqnarray}
  \end{subequations}
in the FF. 

We also report that under a local Weyl re-scaling of the metric, $g'_{\mu\nu}= \mathrm{e}^{-2\phi(x)}g_{\mu\nu}$, the tensor fields $\mathring{O}^N_{\lambda\mu\nu}$ must have conformal weight $-2$, and thus, transform as the metric, whereas 
\begin{equation}
    T'_\mu = T_\mu,\qquad \rho'_\mu = \rho_\mu +2(n+2)\partial_\mu\phi,\qquad u'_\mu=u_\mu-2(n-1)\partial_\mu\phi.
\end{equation}
Clearly, the combination $\rho_\mu + (n+2)u_\mu/(n-1)$ is itself a Weyl invariant. It corresponds to $3(n\Check{Q}_\mu - Q_\mu)/(n-1)$ in the FF. We shall now proceed with a highly pedagogical example.

\section{Revisiting the Hilbert-Palatini action}
\label{sec:HilbertNew}
 \subsection{FF vs. AF}
The $n$-dimensional Hilbert-Palatini (HP) action reads 
\begin{equation}\label{eq:HPactionFF}
    I_{HP}=\frac{1}{2}\int \sqrt{-\mathsf{g}}d^nx R,
\end{equation}
in units $\hbar = c = M_{Pl}=1$ where $M_{Pl}$ is the reduced Planck mass. This is the standard FF action which is invariant under the so-called \emph{projective} transformation 
\begin{equation}\label{eq:LCTprojective}
    \Gamma'{}^\lambda{}_{\mu\nu} = \Gamma^\lambda{}_{\mu\nu}+\frac{1}{n-1}\delta^\lambda_\mu \xi_\nu,
\end{equation}
with $\xi_\mu$ being an arbitrary vector field. 

The field equations in the FF read 
\begin{subequations}\label{eq:HPFFFE}
    \begin{eqnarray}
        2E_{\mu\nu}&\equiv& R_{(\mu\nu)}-\frac{1}{2}Rg_{\mu\nu}=0,\\
        2\Delta_\lambda{}^{\mu\nu}&\equiv&\delta_{\lambda }^{\nu } N^{\mu \alpha }{}_{\alpha } -  N^{\mu \nu }{}_{\lambda } -  N^{\nu }{}_{\lambda }{}^{\mu } + N^{\alpha }{}_{\lambda \alpha } g^{\mu \nu }=0.
    \end{eqnarray}
\end{subequations}
We chose to express the connection field equations in terms of the distortion tensor in order to achieve a more compact output. The invariance of eqs.~\eqref{eq:HPFFFE} under~\eqref{eq:LCTprojective} can be easily seen from the fact that
\begin{equation}
    R'_{\mu\nu}=R_{\mu\nu}+\frac{2}{n-1}\partial_{[\mu}\xi_{\nu]},\qquad \Delta_\lambda{}^{\lambda\mu}\equiv 0,
\end{equation}
the right one holding true identically (off-shell).

It is a well-known fact that the solution to $\Delta_\lambda{}^{\mu\nu}=0$ is the affine connection 
\begin{equation}
    \Gamma^\lambda{}_{\mu\nu}=\Tilde{\Gamma}^\lambda{}_{\mu\nu}+\delta^\lambda_\mu V_\nu,\label{eq:HPConSol}
\end{equation}
where $V_\mu$ is some undetermined vector field. Since 
\begin{equation}
    \Gamma^\lambda{}_{\mu\nu}=\Tilde{\Gamma}^\lambda{}_{\mu\nu}+\delta^\lambda_\mu \left( V_\nu + \frac{1}{n-1}\xi_\nu\right)
\end{equation}
is also a solution, we conclude that the affine connection solving the connection field equations is just the Levi-Civita connection up to the choice of gauge. The effective form of the metric field equations becomes
\begin{equation}
    \Tilde{R}_{\mu\nu}=\frac{1}{2}\Tilde{R}g_{\mu\nu},
\end{equation}
i.e., the HP action is effectively Einstein gravity. 

On the other hand, in the AF, whenever we write $R$ we just mean the expression
\begin{equation}\label{eq:AFscal}
    \tilde{R} + R_{T}+R_{V}+\tilde{\nabla}_\mu\left( 2 T^\mu + u^\mu\right),
\end{equation}
where $\tilde{R}$ is the Riemannian Ricci scalar and 
\begin{subequations}\label{eq:AFscalConstituents}
    \begin{eqnarray}
        R_{T}&=&- \frac{1}{4} H^2 -  \frac{1}{4} \mathring{\pi}{}^2+\frac{1}{2} \mathring{q}{}_{\lambda\mu\nu}\mathring{q}{}^{\lambda\mu\nu}+\frac{1}{2} \mathring{t}{}_{\lambda\mu\nu}\mathring{t}{}^{\lambda\mu\nu} + \mathring{q}{}^{\lambda \nu \mu } \mathring{t}{}_{\nu \mu \lambda },\\
        R_{V}&=& \frac{n-1}{36(n+2)}\rho^2-\frac{n-2}{n-1}T^2+\frac{5-2n}{9(n-1)}u^2+\frac{1}{18}\rho_\mu u^\mu - \frac{n-2}{n-1}T_\mu u^\mu.
    \end{eqnarray}
\end{subequations}
Therefore, up to surface terms, our AF action reads
\begin{equation}\label{eq:HPactionAF}
    I_{HP}=\frac12 \int\sqrt{-\mathsf{g}}d^nx\left(\Tilde{R}+ R_{T}+R_{V}\right).
\end{equation}
The analogue of a projective transformation in the AF is comprised of the simultaneous translations
\begin{equation}\label{eq:AFAtfs}
    A'{}^I = A^I + a^{I-1},
\end{equation}
with 
\begin{equation}
    \xi_\mu\equiv a^{\hz}_\mu=\frac{n-1}{2(n+2)}a^{\ho}_\mu = -\frac12 a^{\htw}_\mu.
\end{equation}
One can easily verify that the above transformations should only affect $R_V$. Since it happens that $R_V$ is invariant, the transformations~\eqref{eq:AFAtfs} constitute a symmetry of the full action~\eqref{eq:HPactionAF}. 

Now, there are two equivalent ways to proceed as we have shown in the previous section. We can either use eqs.~\eqref{eq:HPFFFE} to reconstruct the field equations in the AF, or we can directly vary the integral~\eqref{eq:HPactionAF} with respect to the AF field variables (quickest strategy). Both methods lead to the same result, namely the field equations
\begin{subequations}\label{eq:HPAFFEcon}
    \begin{eqnarray}
        \mathring{\mathcal{O}}{}^{\lambda\mu\nu}_{\hz}&\equiv&-\frac14 H^{\lambda\mu\nu}=0,\qquad
        \mathring{\mathcal{O}}{}^{\lambda\mu\nu}_{\ho}\equiv\frac{1}{2}\left( \mathring{t}{}^{\lambda\mu\nu}-\mathring{q}{}^{[\mu\nu]\lambda}\right)=0,\\
        \mathring{\mathcal O}{}^{\lambda\mu\nu}_{\htw}&\equiv&-\frac14 \mathring{\pi}{}^{\lambda\mu\nu}=0,\qquad
        \mathring{\mathcal O}{}^{\lambda\mu\nu}_{\hth}\equiv\mathring{\mathcal{O}}{}^{(\mu\nu)\lambda}_{\ho}+\frac{1}{8}\mathring{q}^{\lambda\mu\nu}=0,\\
        {\mathcal{A}}^\mu_{\ho}&\equiv&-\frac{n-2}{n-1}\left( T^\mu+\frac12 u^\mu\right)=0,\\
        \mathcal{A}^\mu_{\htw} &\equiv& \frac{n-1}{36(n+2)}\left( \rho^\mu + \frac{n+2}{n-1} u^\mu \right)=0,\\
        \mathcal{A}^\mu_{\hth} &\equiv& \frac{1}{2}\mathcal{A}^\mu_{\ho}+\frac{n+2}{n-1}\mathcal{A}^\mu_{\htw}=0,
    \end{eqnarray}
\end{subequations}
and 
\begin{eqnarray}\label{eq:HPAFFEmet}
    2\hat{E}_{\mu\nu}&\equiv&\Tilde{R}_{\mu\nu}-\frac{1}{2}g_{\mu\nu}\left(\Tilde{R}+R_{T}+R_{V}\right)-\frac{3}{4}\left( H_{\mu\alpha\beta}H_{\nu}{}^{\alpha\beta}+\mathring{\pi}{}_{\mu\alpha\beta}\mathring{\pi}{}_{\nu}{}^{\alpha\beta}\right)+\nonumber\\
    &&+\mathring{q}{}_{\alpha\beta\mu}\mathring{q}{}^{\alpha\beta}{}_\nu+\mathring{t}{}_{\alpha\beta\mu}\mathring{t}{}^{\alpha\beta}{}_\nu+\frac12\left(\mathring{q}{}_{\mu\alpha\beta}\mathring{q}{}_{\nu}{}^{\alpha\beta}+\mathring{t}{}_{\mu\alpha\beta}\mathring{t}{}_{\nu}{}^{\alpha\beta}\right)+\nonumber\\
    &&+\mathring{t}{}_{\alpha\beta(\mu}\mathring{q}{}_{\nu)}{}^{\alpha\beta}-\mathring{q}{}^{\alpha\beta}{}_{(\mu}\mathring{t}{}_{\nu)\alpha\beta}-\mathring{t}{}_{\beta\alpha(\mu} \mathring{q}{}^{\alpha\beta}{}_{\nu)}-\nonumber\\
    &&-\frac{n-2}{n-1}T_\mu T_\nu + \frac{n-1}{36(n+2)}\rho_\mu\rho_\nu+\frac{5-2n}{9(n-1)}u_\mu u_\nu -\nonumber\\
    &&-\frac{n-2}{n-1}T_{(\mu}u_{\nu)}+\frac{1}{18}\rho_{(\mu} u_{\nu)}=0.
\end{eqnarray}

It is evident that the first two lines in~\eqref{eq:HPAFFEcon} suggest that $\mathring{O}^N_{\lambda\mu\nu}=0$. The remaining two independent equations, $\mathcal{A}^\mu_{\ho}=0=\mathcal{A}^\mu_{\htw}$, do further imply that 
\begin{equation}
   u_\mu=-2T_\mu=-\frac{n-1}{n+2}\rho^\mu.
\end{equation}
Hence, the full solution to the system~\eqref{eq:HPAFFEcon} is 
\begin{equation}
    \mathring{O}^N_{\lambda\mu\nu}=0,\qquad V_\mu\equiv u_\mu=-2T_\mu=-\frac{n-1}{n+2}\rho^\mu,
\end{equation}
where $V_\mu$ is an arbitrary vector field. Since 
\begin{equation}
    T_\mu=-\frac{1}{2}V_\mu + \xi_\mu,\qquad \rho_\mu=\frac{n+2}{n-1}\left(2\xi_\mu -V_\mu\right),\qquad u_\mu = V_\mu-2\xi_\mu
\end{equation}
is also a solution, we conclude that $A^I_\mu=0$ up to the choice of gauge. The effective form of the metric field equations~\eqref{eq:HPAFFEmet} becomes
\begin{equation}
    \Tilde{R}_{\mu\nu}=\frac{1}{2}\Tilde{R}g_{\mu\nu},
\end{equation}
i.e., the HP action in the AF is again, in effect, Einstein gravity. 

\subsection{Projective symmetry and the AF$^{\circ}$}

The careful reader would have already noticed that what is a projective symmetry in the FF manifests itself as a true gauge symmetry in the AF. Indeed, eqs.~\eqref{eq:HPAFFEcon} reveal that there are only two independent equations for the triplet $A_\mu$ which rather signals that one of these field variables is after all redundant. First, bear in mind that the combinations
\begin{equation}
    T_\mu+\frac{1}{2}u_\mu,\qquad  \rho^\mu + \frac{n+2}{n-1} u^\mu,
\end{equation}
which are invariant under~\eqref{eq:AFAtfs}, correspond to the FF combinations
\begin{equation}
    \frac{1}{2} \left(\check{Q}_{\mu } -  Q_{\mu }\right) + T_{\mu },\qquad \frac{3}{n-1}\left(n \Check{Q}_\mu - Q_\mu\right),
\end{equation}
respectively, which are invariant under~\eqref{eq:LCTprojective}.

Let us then discuss the idea that for $n>2$, whenever the projective symmetry is at play, one should favor a doublet 
\begin{equation}
    B_\mu = \left\lbrace \frac23\left(T_\mu+\frac{1}{2}u_\mu\right), \frac{n-1}{9\sqrt{n^2-4}}\left(\rho^\mu + \frac{n+2}{n-1} u^\mu\right)\right\rbrace\label{def:DAFBdoublet}
\end{equation}
over the redundant triplet $A_\mu$. Note that the above choice of $B^A$, where indices $A,B,...$ assume values in $\{\hz,\ho \}$, is not the most general one. Nevertheless, it is the most convenient choice for our purposes here since it casts $R_V$ into the neat form
\begin{equation}\label{eq:DAFRV}
    \mathcal{R}_V = \frac{9(n-2)}{4(n-1)}\eta_{AB}B^A_\mu B^B_\nu g^{\mu\nu},
\end{equation}
where $\eta_{AB}$ are the components of the two-dimensional Minkowski metric $\eta^{(2)}=\mathrm{diag}(-1,1)$. 

The affected parts of~\eqref{eq:HPAFFEcon} read 
\begin{equation}
    \mathcal{A}^\mu_{\ho}\equiv -\frac{3(n-2)}{2(n-1)}B^\mu_{\hz}=0,\qquad \mathcal{A}^\mu_{\htw}\equiv \frac{1}{4}\sqrt{\frac{n-2}{n+2}}B^\mu_{\ho}=0,
\end{equation}
whereas the metric field equations~\eqref{eq:HPAFFEmet} are rendered into 
\begin{eqnarray}
    2\hat{E}_{\mu\nu}&\equiv&\Tilde{R}_{\mu\nu}-\frac{1}{2}g_{\mu\nu}\left(\Tilde{R}+R_{T}+\mathcal{R}_{V}\right)-\frac{3}{4}\left( H_{\mu\alpha\beta}H_{\nu}{}^{\alpha\beta}+\mathring{\pi}{}_{\mu\alpha\beta}\mathring{\pi}{}_{\nu}{}^{\alpha\beta}\right)+\nonumber\\
    &&+\mathring{q}{}_{\alpha\beta\mu}\mathring{q}{}^{\alpha\beta}{}_\nu+\mathring{t}{}_{\alpha\beta\mu}\mathring{t}{}^{\alpha\beta}{}_\nu+\frac12\left(\mathring{q}{}_{\mu\alpha\beta}\mathring{q}{}_{\nu}{}^{\alpha\beta}+\mathring{t}{}_{\mu\alpha\beta}\mathring{t}{}_{\nu}{}^{\alpha\beta}\right)+\nonumber\\
    &&+\mathring{t}{}_{\alpha\beta(\mu}\mathring{q}{}_{\nu)}{}^{\alpha\beta}-\mathring{q}{}^{\alpha\beta}{}_{(\mu}\mathring{t}{}_{\nu)\alpha\beta}-\mathring{t}{}_{\beta\alpha(\mu} \mathring{q}{}^{\alpha\beta}{}_{\nu)}+\frac{9(n-2)}{4(n-1)}\eta_{AB}B^A_\mu B^B_\nu=0.
\end{eqnarray}
Interestingly, when written in terms of the $B^A$ fields, the HP action and its field equations exhibit an $SO(1,1)$ symmetry! Indeed, the field transformation $B'_\mu = \Lambda(x) B_\mu$ with 
\begin{equation}
    \Lambda = \begin{pmatrix} 
    \cosh\theta(x) & \sinh\theta(x) \\ 
    \sinh\theta(x) & \cosh\theta(x)
    \end{pmatrix},
\end{equation}
preserves both of them. We remark that the manifestation of this transformation as a group action on the field variables is exclusive to the use of the $B^A$'s to formulate the HP action. 

Since the use of these field re-combinations revealed something new, we find it worth to take a step back and generalize the whole thing to another framework which we dub AF$^{\circ}$ or ``diminished alternative framework''. In the AF$^{\circ}$, we formulate our MAG theory in terms of the reduced set $\{g,\mathring{O}^N,B^A\}$ of field variables. The doublet $B_\mu$ should consist of two linear combinations of AF vector fields which are invariant under~\eqref{eq:AFAtfs}. Equivalently, it should be comprised of two linear combinations of $T_\mu,Q_\mu,\Check{Q}_\mu$ which are invariant under~\eqref{eq:LCTprojective}, the point being that in the AF$^{\circ}$ such transformations should constitute an identity operation on our field variables. The most general combinations invariant under~\eqref{eq:AFAtfs} are 
\begin{equation}
    B^{\hz}_\mu = \alpha_{I-1} A^I_\mu,\qquad B^{\ho}_\mu = \beta_{I-1} A^I_\mu,\label{def:DAFBdoubletGeneral}
\end{equation}
with
\begin{equation}
    \alpha_{\htw}=\frac{\alpha_{\hz}}{2}+\frac{(n+2)\alpha_{\ho}}{n-1},\label{eq:DAFCoefTune}
\end{equation}
and ditto for the coefficients $\beta_{A}$.

Let
\begin{equation}
    \mathcal{B}^\mu_A = \frac{1}{\sqrt{-\mathsf{g}}}\frac{\delta I}{\delta B^A_\mu},
\end{equation}
such that the field equations for the field $B^A$ are $\mathcal{B}^\mu_A=0$. We can directly make the identifications 
    \begin{eqnarray}
        \mathcal{A}^\mu_{I}&=&\alpha_{I-1}\mathcal{B}^\mu_{\hz}+\beta_{I-1}\mathcal{B}^\mu_{\ho},\label{eq:DAFtoAF}
    \end{eqnarray}
where one has to remember that the coefficients obey the relation~\eqref{eq:DAFCoefTune}. Clearly, whenever $\mathcal{B}^\mu_A=0$, it follows that $\mathcal{A}^\mu_I=0$. However, whenever $\mathcal{A}^\mu_I=0$, it follows that $\mathcal{B}^\mu_A=0$ only when
\begin{equation}
    \alpha_{\hz}\beta_{\ho}-\alpha_{\ho}\beta_{\hz}\neq 0.\label{eq:DAFEquivCondition}
\end{equation}
Therefore, the field equations in the AF$^{\circ}$ imply and are, under assumptions, implied by the field equations in the AF or in the FF (if we follow the equivalence chain).

Lastly, let us see exactly how we ended up with~\eqref{def:DAFBdoublet}. In terms of the fields $B^A$, as defined in eq.~\eqref{def:DAFBdoubletGeneral}, we have that 
\begin{equation}
    R_V = \left[f(\beta_{\hz}^2,\beta_{\ho}^2)B^{\hz}_\mu B^{\hz}_\nu - 2 f(\alpha_{\hz}\beta_{\hz},\alpha_{\ho}\beta_{\ho})B^{\hz}_\mu B^{\ho}_\nu+f(\alpha_{\hz}^2,\alpha_{\ho}^2)B^{\ho}_\mu B^{\ho}_\nu\right]g^{\mu\nu},
\end{equation}
where 
\begin{equation}
    f(x,y):=\frac{(n-1)^2x-36(n^2-4)y}{36(n^2+n-2)(\alpha_{\ho}\beta_{\hz}-\alpha_{\hz}\beta_{\ho})^2}.
\end{equation}
Moreover, the total divergence in~\eqref{eq:AFscal} assumes the form 
\begin{equation}
    \frac{2}{\alpha_{\ho}\beta_{\hz}-\alpha_{\hz}\beta_{\ho}}\Tilde{\nabla}_\mu\left( \alpha_{\ho}B^\mu_{\ho}-\beta_{\ho}B^\mu_{\hz}\right).
\end{equation}
To get the above, we expressed $\alpha_{\htw}$ in terms of $\alpha_{\hz},\alpha_{\ho}$ via eq.~\eqref{eq:DAFCoefTune}, and ditto for the parameters $\beta_{A}$. Different choices for the parameters $\alpha_A,\beta_A$ obviously amount to different changes of field variables. 

A convenient choice is one for which $f(\alpha_{\hz}\beta_{\hz},\alpha_{\ho}\beta_{\ho})=0$, namely
\begin{equation}
    \beta_{\ho}=\frac{(n-1)^2\alpha_{\hz}\beta_{\hz}}{36(n^2-4)\alpha_{\ho}},
\end{equation}
provided $n>2$, which yields
\begin{equation}
    R_V = \frac{(n-2)(n-1)}{(n-1)^2\alpha_{\hz}^2 - 36(n^2-4)\alpha_{\ho}^2}\left[ - B^{\hz}_\mu B^{\hz}_\nu+\frac{36\alpha_{\ho}^2(n^2-4)}{\beta_{\hz}^2(n-1)^2} B^{\ho}_\mu B^{\ho}_\nu\right]g^{\mu\nu}.
\end{equation}
Further imposing that 
\begin{equation}
    \beta_{\hz}=\frac{6|\alpha_{\ho}|\sqrt{n^2-4}}{n-1},
\end{equation}
gives
\begin{equation}
    R_V = \frac{(n-2)(n-1)}{(n-1)^2\alpha_{\hz}^2 - 36(n^2-4)\alpha_{\ho}^2}\eta_{AB}B^A_\mu B^B_\nu g^{\mu\nu}.
\end{equation}
Finally, we choose 
\begin{equation}
    \alpha_{\ho} = \frac{n-1}{18}\sqrt{\frac{9\alpha_{\hz}^2-4}{n^2-4}},
\end{equation}
for later convenience, which leads to
\begin{equation}
    {R}_V = \frac{9(n-2)}{4(n-1)}\eta_{AB}B^A_\mu B^B_\nu g^{\mu\nu}=:\mathcal{R}_V.
\end{equation}

Note that all of the above parameter choices are in agreement with~\eqref{eq:DAFEquivCondition} which becomes 
\begin{equation}
    \frac{2(n-1)}{27\sqrt{n^2}-4}\neq 0.
\end{equation}
In terms of the AF fields, our new field variables, $B^A$, read 
\begin{subequations}
    \begin{eqnarray}
        B^{\hz}_\mu &=& \alpha_{\hz} T_\mu + \frac{n-1}{18}\sqrt{\frac{9\alpha_{\hz}^2-4}{n^2-4}}\rho_\mu+\frac{1}{18}\left( 9 \alpha_{\hz}+\sqrt{\frac{(9\alpha_{\hz}^2-4)(n+2)}{n-2}}\right) u_\mu,\\
        B^{\ho}_\mu &=& \frac{\sqrt{9\alpha_{\hz}^2-4}}{3} T_\mu + \frac{(n-1)\alpha_{\hz}}{6\sqrt{n^2-4}}\rho_\mu+\frac{1}{6}\left( \sqrt{\frac{n+2}{n-2}}\alpha_{\hz}+\sqrt{9\alpha_{\hz}^2-4}\right) u_\mu,
    \end{eqnarray}
\end{subequations}
where we may further fix $|\alpha_{\hz}|=2/3$ so that $B^{\ho}$ is purely a combination of traces of the non-metricity tensor. This brings us to~\eqref{def:DAFBdoublet}. Henceforth, the word AF$^{\circ}$ will always mean that we use the specific doublet~\eqref{def:DAFBdoublet}.

\subsection{$SO(1,2)$ symmetry, torsion/non-metricity rotations, and the d-AF$^{\circ}$}

Now, we restrict ourselves to $n=4$ space-time dimensions where things get a bit more interesting. Via the dualization~\eqref{def:PseudoVector}, we have a pseudo-vector, and the term $\propto H^2$ can be moved from $R_{T}$ to $R_{V}$. In particular, let us introduce the objects
\begin{subequations}\label{eq:DAFscalConstituents}
    \begin{eqnarray}
        \hat{R}_{T}&=& -  \frac{1}{4} \mathring{\pi}{}^2+\frac{1}{2} \mathring{q}{}_{\lambda\mu\nu}\mathring{q}{}^{\lambda\mu\nu}+\frac{1}{2} \mathring{t}{}_{\lambda\mu\nu}\mathring{t}{}^{\lambda\mu\nu} + \mathring{q}{}^{\lambda \nu \mu } \mathring{t}{}_{\nu \mu \lambda },\\
        \hat{\mathcal{R}}_{V}&=&\frac{3}{2}\eta_{\mathcal{A}\mathcal{B}}B^{\mathcal{A}}_\mu B^{\mathcal{B}}_\nu g^{\mu\nu}.
    \end{eqnarray}
\end{subequations}
where the calligraphic indices take values in $\{ \hz,\ho,\htw\}$, $\eta_{\mathcal{A}\mathcal{B}}$ are the components of the three-dimensional Minkowski metric, $\eta^{(3)}=\mathrm{diag}(-1,1,1)$, and we have formed a triplet
\begin{equation}\label{def:DAFBtriplet}
    B_\mu{} = \left\lbrace B_\mu^{\hz},B_\mu^{\ho},S_\mu\right\rbrace,
\end{equation}
with $B^A$ given by~\eqref{def:DAFBdoublet}.

The affected parts of~\eqref{eq:HPAFFEcon} read 
\begin{equation}
    \mathcal{A}^\mu_{\ho}\equiv -\frac{3(n-2)}{2(n-1)}B^\mu_{\hz}=0,\qquad \mathcal{A}^\mu_{\htw}\equiv \frac{1}{4}\sqrt{\frac{n-2}{n+2}}B^\mu_{\ho}=0,
\end{equation}
and 
\begin{equation}
    \mathring{\mathcal{O}}^{\lambda\mu\nu}_{\hz}\equiv \frac{1}{4}\Tilde{\epsilon}^{\lambda\mu\nu\alpha}S_\alpha=0,
\end{equation}
whereas the metric field equations~\eqref{eq:HPAFFEmet} are rendered into 
\begin{eqnarray}
    2\hat{E}_{\mu\nu}&\equiv&\Tilde{R}_{\mu\nu}-\frac{1}{2}g_{\mu\nu}\left(\Tilde{R}+\hat{R}_{T}+\hat{\mathcal{R}}_{V}\right)-\frac{3}{4}\mathring{\pi}{}_{\mu\alpha\beta}\mathring{\pi}{}_{\nu}{}^{\alpha\beta}+\nonumber\\
    &&+\mathring{q}{}_{\alpha\beta\mu}\mathring{q}{}^{\alpha\beta}{}_\nu+\mathring{t}{}_{\alpha\beta\mu}\mathring{t}{}^{\alpha\beta}{}_\nu+\frac12\left(\mathring{q}{}_{\mu\alpha\beta}\mathring{q}{}_{\nu}{}^{\alpha\beta}+\mathring{t}{}_{\mu\alpha\beta}\mathring{t}{}_{\nu}{}^{\alpha\beta}\right)+\nonumber\\
    &&+\mathring{t}{}_{\alpha\beta(\mu}\mathring{q}{}_{\nu)}{}^{\alpha\beta}-\mathring{q}{}^{\alpha\beta}{}_{(\mu}\mathring{t}{}_{\nu)\alpha\beta}-\mathring{t}{}_{\beta\alpha(\mu} \mathring{q}{}^{\alpha\beta}{}_{\nu)}+\frac32 \eta_{\mathcal A\mathcal B}B^{\mathcal A}_\mu B^{\mathcal{B}}_\nu=0.
\end{eqnarray}
Remarkably, when written in terms of the triplet~\eqref{def:DAFBtriplet}, the four-dimensional HP action and its field equations exhibit a larger symmetry under an $SO(1,2)$ group action mixing the components $B^{\mathcal{A}}_\mu$.

Of particular interest is the transformation $B'_\mu = \Lambda(x) B_\mu$ with
\begin{equation}\label{eq:SO(2)symmetrySub}
    \Lambda=\begin{pmatrix}
        1&&\\&\cos \theta(x) & \sin \theta(x)\\&-\sin\theta(x)&\cos\theta(x)
    \end{pmatrix},
\end{equation}
which represents an $SO(2)$ rotation in the $\{B^{\ho}_\mu,S_\mu\}$ (field) subspace. Specifically, the invariance of the action under the discrete transformation with $\Lambda(\theta=\pi/2)$, constitutes an exceptional example of a symmetry under \emph{torsion/non-metricity rotations}. Indeed, omitting the space-time indices, we have
\begin{equation}
    \begin{pmatrix}
        B^{\hz}\\ B^{\ho}\\ S
    \end{pmatrix} \to \begin{pmatrix}
        B^{\hz}\\ S\\ -B^{\ho}
    \end{pmatrix},
\end{equation}
where $S_\mu$ is pure torsion (pseudo-vector) and $B^{\ho}_\mu$ is pure non-metricity (traces), as defined in~\eqref{def:PseudoVector} and~\eqref{def:DAFBdoublet}, respectively. Do further note that if we consider $B^{\ho}_\mu$ and $S_\mu$ as, respectively, the real and imaginary parts of a complex vector
\begin{equation}\label{def:RSV}
    \tau_\mu = B^{\ho}_\mu + i S_\mu,
\end{equation}
we have that 
\begin{equation}
    \hat{\mathcal{R}}_V=-\frac{3}{2}\left( B^{\hz}_\mu B^{\hz}_\nu - \tau_\mu \tau^*_\nu\right)g^{\mu\nu},
\end{equation}
where ${\tau}^*$ is the complex conjugate of $\tau$. The previous $SO(2)$ symmetry now manifests itself as a $U(1)$ under $\tau'_\mu = \mathrm{e}^{-i\theta(x)} \tau_\mu$. 

As a closing remark, let us mention here that in what follows we will often use the torsion pseudo-vector $S_\mu$ instead of $H_{\lambda\mu\nu}$ as a field variable in our four-dimensional projective-invariant examples. In other words, we will be using the set $\{ g, \mathring{O}^I, B^{\mathcal{A}}\}$ when invoking the AF$^{\circ}$, and some clarifications are in order. Letting an action
\begin{equation}
    I=\int\sqrt{-\mathsf{g}}d^4x\mathcal{L}[g,\mathring{O}{}^{I},\mathcal{B}^{\mathcal{A}}]
\end{equation}
vary, we get 
\begin{equation}\label{eq:dDAFvariation}
    \delta I = \int \sqrt{-\mathsf{g}}d^4x \left( \check{E}_{\mu\nu}\delta g^{\mu\nu} +\mathring{O}{}^{\lambda\mu\nu}_{I} \delta \mathring{O}{}_{\lambda\mu\nu}^{I} + \mathcal{B}^\mu_{\mathcal{A}}\delta B_\mu^{\mathcal{A}} \right) + \text{s.t.}.
\end{equation}
Therefore, the field equations read 
\begin{equation}
    \Check{E}_{\mu\nu}=0,\qquad \mathring{\mathcal{O}}^{\lambda\mu\nu}_{I}= 0,\qquad {\mathcal{B}}^\mu_{\mathcal{A}}= 0,
\end{equation}
where $\Check{E}_{\mu\nu}$ is a symmetric tensor. Following the preceded steps, one should be able to directly make the identifications
\begin{subequations}
\begin{eqnarray}
        \mathcal{B}^\mu_{\hz} &=& \frac32 \mathcal{A}^\mu_{\ho},\qquad
        {\mathcal{B}}^\mu_{\ho} = 6\sqrt{3}\mathcal{A}^\mu_{\htw},\qquad
        \mathcal{B}^\mu_{\htw} = \mathring{\mathcal{O}}{}_{\alpha\beta\gamma}^{\hz}\tilde{\epsilon}^{\mu\alpha\beta\gamma},\\
        \mathcal{A}^\mu_{\hth} &=& \frac13\left( \mathcal{B}^\mu_{\hz} + \frac{1}{\sqrt{3}}\mathcal{B}^\mu_{\ho}\right),\qquad
        \Check{E}_{\mu\nu} = \hat{E}_{\mu\nu}+\mathcal{B}_{(\mu}^{\htw}S_{\nu)}-\frac{1}{2}g_{\mu\nu}S_\alpha \mathcal{B}^\alpha_{\htw},
\end{eqnarray}
\end{subequations}
which tell us how the functional derivatives in the two frameworks are related. These do once again suffice to prove the equivalence between the field equations in the AF and in this framework which, for the sake of clarity and in lack of a better name, we call d-AF$^{\circ}$, with the d reminding us that we use the dual of $H_{\lambda\mu\nu}$. 

\section{Exciting the connection: a series of examples}
\label{sec:Excited}
It has been standard practice in the MAG community to motivate actions from a geometric perspective and with full generality in mind. Although this is in general good practice, it often leads to an intractable set of field equations; in the end one will unavoidably sacrifice generality to get results. In this section we propose an different motivation for MAG models. Using the alternative frameworks we presented, we write meaningful field theories propagating some of the new field variables. These are, in essence, MAG theories propagating certain connection DoF in a tractable and controllable manner. 

These MAG theories are inspired by Riemannian theories with additional fields. In fact, they yield an effective\footnote{The word ``effective'' is used to denote that one has eliminated auxiliary field variables, i.e., fields which do not appear with time derivatives in the field equations.} set of field equations which is very much identical to the corresponding system of equations in the Riemannian case, the crucial difference being that instead of additional fields we use specific modes of the distortion tensor. The obvious advantage is that for a given symmetry ansatz, we exactly know how to solve the differential equations that arise. Hence, one should not expect metric solutions novel in form; the only novelty is that these known metric backgrounds are now part of a larger solution with torsion and non-metricity. In what follows, we will occasionally omit space-time indices (or internal indices) when they are trivially understood. 

\subsection{\boldmath The MAGswell theory}
The action for a Maxwell field $A_\mu$ coupled to four-dimensional gravity with a cosmological constant is 
  \begin{equation}\label{eq:EMaction}
    \tilde{I}_{EM}=\frac12 \int\sqrt{-\mathsf{g}}d^4x \left(\tilde{R}-2\Lambda-\frac{1}{2}F^2\right),
  \end{equation}
  where $F_{\mu\nu}=2\partial_{[\mu}A_{\nu]}$ is the field-strength tensor. We remind the reader that we work in natural units with the reduced Planck mass further set to unity. The above integral is invariant under shifts $A' = A+d\theta$, where $\theta(x)$ is some scalar potential. Variation with respect to the metric yields the metric field equations
    \begin{eqnarray}
      \tilde{G}_{\mu\nu}+\Lambda g_{\mu\nu}&=&F_{\mu}{}^\alpha F_{\nu\alpha}-\frac{1}{4}F^2g_{\mu\nu},\label{eq:EMmeteom}
    \end{eqnarray}
where $\tilde{G}_{\mu\nu}=\tilde{R}_{\mu\nu}-\tfrac{1}{2}\tilde{R}g_{\mu\nu}$ is the Einstein tensor. The Maxwell equations and the Bianchi identity can be written as
\begin{equation}
  \partial_{\nu}\left(\sqrt{-g}F^{\nu\mu}\right)=0,\qquad \partial_{\nu}\left(\sqrt{-g}\ast F^{\nu\mu}\right)=0,\label{eq:EMFeom}
\end{equation}
respectively, where $\ast F^{\mu\nu}=\tfrac12 \tilde{\epsilon}{}^{\mu\nu\rho\lambda}F_{\rho\lambda}$ is the Hodge dual of the field-strength tensor. 

Here, we wish to write down a more or less similar theory for a massless vector in MAG, the homophonous \emph{MAGswell field}, $C_\mu$, as we may playfully dub it. We will do so by considering the Ricci scalar as our cornerstone and adding proper terms to it. Since the HP action is invariant under projective transformations we would like to retain this feature in the complete action. This will also allow us to work in the AF$^{\circ}$. We emphasize that the MAGswell field should be understood as part of the geometry of space-time, i.e., it is not the familiar gauge connection.

Without further ado, let us consider a fairly general projective-invariant candidate for the four-dimensional MAGswell action which in the d-AF$^{\circ}$ assumes the form
\begin{equation}\label{eq:MAGswellAction}
    I_C=\int \sqrt{-\mathsf{g}}d^4x \mathcal{L}_{{C}}\equiv \int \sqrt{-\mathsf{g}}d^4x\left(\mathcal{L}_{{HP}}+\mathcal{L}_{ct}+\mathcal{L}_{kin}\right),
\end{equation}
with
\begin{subequations}\label{eq:MAGswellConstituents}
\begin{eqnarray}
    \mathcal{L}_{{HP}} &=& \frac{1}{2}\left(R-2\Lambda\right),\\
    \mathcal{L}_{ct} &=& -\frac{3}{4}g^{\mu\nu}B_\mu^{A}B_\nu^{B}\eta_{AB},\label{def:CounterLag}\\
    \mathcal{L}_{kin} &=& -\frac{1}{4}F_{(C)}^2.\label{def:KinLag}
\end{eqnarray}
\end{subequations}
Here, ${F}^{(C)}_{\mu\nu}=2\partial_{[\mu}C_{\nu]}$ with $C_\mu\equiv \alpha_A B^{A}_\mu$ being the composite MAGswell field and $\alpha_A$ dimensionless constants, i.e., real non-zero numbers. The curvature scalar $R$ stands for
\begin{equation}
    \Tilde{R}+\hat{R}_T+\hat{\mathcal{R}}_V+3\Tilde{\nabla}_\mu B^{\mu}_{\hz}
\end{equation}
with the constituents given in~\eqref{eq:DAFscalConstituents}. The form of the action in the AF, or the FF, can be easily obtained by remembering that 
\begin{subequations}\label{eq:DAFtoAFtoFF}
    \begin{eqnarray}
        B_\mu^{\hz} &=& \frac{1}{3}\left( 2 T_\mu + u_\mu\right) = \frac{1}{3} \left(\check{Q}_{\mu } -  Q_{\mu } + 2 T_{\mu }\right),\\
        B_\mu^{\ho} &=& \frac{1}{6 \sqrt{3}}\left(\rho_{\mu } + 2 u_{\mu }\right)=\frac{1}{6 \sqrt{3}}\left(4 \check{Q}_{\mu } -  Q_{\mu }\right).
    \end{eqnarray}
\end{subequations}
For example, in the FF we have 
    \begin{eqnarray}
        \mathcal{L}_{ct} &=& - \frac{1}{36} \check{Q}_{\mu } \check{Q}^{\mu } -  \frac{1}{9} \check{Q}^{\mu } Q_{\mu } + \frac{11}{144} Q_{\mu } Q^{\mu } + \frac{1}{3} \left( \check{Q}^{\mu } T_{\mu } -   Q^{\mu } T_{\mu } +  T_{\mu } T^{\mu }\right),
    \end{eqnarray}
and $\mathcal{L}_{kin}$ is a linear combination of $F_{(T)}^2$, $F_{(Q)}^2$, $F_{(\Check{Q})}^2$, $F^{(T)}_{\mu\nu}F_{(Q)}^{\mu\nu}$, $F^{(T)}_{\mu\nu} F_{(\check{Q})}^{\mu\nu}$ and $F^{(Q)}_{\mu\nu} F_{(\Check{Q})}^{\mu\nu}$, where 
\begin{equation}
    F^{(T)}_{\mu\nu} = 2\partial_{[\mu}T_{\nu]},\qquad F^{(Q)}_{\mu\nu} = 2\partial_{[\mu}Q_{\nu]}, \qquad F^{(\Check{Q})}_{\mu\nu} = 2\partial_{[\mu}\Check{Q}_{\nu]}.
\end{equation}
A quick inspection of the action indicates that the role of $\mathcal{L}_{ct}$ is that of a counter-term Lagrangian introduced to eradicate the mass terms for the $B^A_\mu$'s, entering via the Ricci scalar, in order for $C_\mu$ to appear massless. 

It is worth noting here that the use of $R$ instead of $\tilde{R}$ in the action~\eqref{eq:MAGswellAction} seems to be in conflict with the whole logic behind the alternative formulation. Indeed, one could use $\Tilde{R}$ and completely avoid the need for the counter-term Lagrangian~\eqref{def:CounterLag}. As it will become clear in what follows, this would lead to another theory whose FF version would possess a much larger gauge freedom, but in which (the theory), the connection solution would again be equivalent to the one we present below up, of course, to the choice of gauge. In this sense, the use of $R$ simply serves as a convenient and very specific partial gauge-fixing mechanism which is by no means a necessity. For example, one could achieve the same result by replacing $R$ with $\Tilde{R}$ in $\mathcal{L}_{HP}$ and by trading~\eqref{def:CounterLag} for a Lagrangian with mass terms for the torsion/non-metricity modes not appearing in~\eqref{def:KinLag}.

\paragraph{Symmetries in the AF$^{\circ}$ (or d-AF$^{\circ}$).} The action~\eqref{eq:MAGswellAction} is invariant under the simultaneous shifts 
\begin{equation}\label{eq:DAFgaugeMAGswellShift}
B'^A = B^A + b^A
\end{equation}
with
\begin{equation}
     b^{\ho}_\mu=\frac{\partial_\mu\theta - \alpha_{\hz}b^{\hz}_\mu}{\alpha_{\ho}},\label{eq:DAFgaugeMAGswell}
\end{equation}
where $\theta(x)$ is some real scalar potential. The number of free gauge parameters to be fixed is thus two.

\paragraph{Symmetries in the AF.} The action~\eqref{eq:MAGswellAction} in the AF is invariant under the simultaneous shifts 
\begin{equation}
    A'^I = A^I + b^{I-1}
\end{equation}
with
\begin{equation}
    b^{\htw}_\mu=-\frac{12\alpha_{\hz}b^{\hz}_\mu+\alpha_{\ho}\sqrt{3}b^{\ho}_\mu-18\partial_\mu\theta}{6\alpha_{\hz}+2\sqrt{3}\alpha_{\ho}}\label{eq:AFgaugeMAGswell}
\end{equation}
The number of free gauge parameters to be fixed is three. 

\paragraph{Symmetries in the FF.} The action~\eqref{eq:MAGswellAction} in the FF is invariant under the connection transformation
\begin{equation}\label{eq:FFgaugeMAGswellShift}
    \Gamma'^\lambda{}_{\mu\nu}=\Gamma^\lambda{}_{\mu\nu} + a^\lambda g_{\mu\nu} + \delta^\lambda_\nu b_\mu + \delta^\lambda_\mu c_\nu ,
\end{equation}
with
\begin{equation}
    b_\mu = \frac{(\alpha_{\hz}+\alpha_{\ho}\sqrt{3})a_\mu - \partial_\mu\theta}{\alpha_{\hz}-\alpha_{\ho}\sqrt{3}}. \label{eq:FFgaugeMAGswell}
\end{equation}
The number of free gauge parameters to be fixed is three. 

Indeed, if $\verb|#|_{fgp}(-)$ outputs the number of free gauge parameters in a certain framework, then $\verb|#|_{fgp}(\text{FF})=\verb|#|_{fgp}(\text{AF})$ in all cases. The fact that $\verb|#|_{fgp}(\text{AF$^{\circ}$})=\verb|#|_{fgp}(\text{AF})-1$ has to do with the projective-symmetry ``charge'' being initially absorbed into the field variables of the AF$^{\circ}$. Having discussed the symmetries in the different frameworks, we now turn our attention to the field equations. 

In the d-AF$^{\circ}$, they read
\begin{subequations}\label{eq:MAGswelldDAFFE}
    \begin{eqnarray}
        \mathring{\mathcal{O}}{}^{\lambda\mu\nu}_{\ho}&\equiv&\frac{1}{2}\left( \mathring{t}{}^{\lambda\mu\nu}-\mathring{q}{}^{[\mu\nu]\lambda}\right)=0,\qquad
        \mathring{\mathcal{O}}{}^{\lambda\mu\nu}_{\htw}\equiv-\frac{1}{4} \mathring{\pi}{}^{\lambda\mu\nu}=0,\nonumber\\
        \mathring{\mathcal O}{}^{\lambda\mu\nu}_{\hth}&\equiv&\mathring{\mathcal{O}}{}^{(\mu\nu)\lambda}_{\ho}+\frac{1}{8}\mathring{q}^{\lambda\mu\nu}=0,\\
        \mathcal{B}^\mu_{\hz}&\equiv&\frac{\alpha_{\hz}}{\alpha_{\ho}}\mathcal{B}^\mu_{\ho}\equiv\alpha_{\hz} \Tilde{\nabla}_\nu{F}_{(C)}^{\nu\mu}=0,\qquad \mathcal{B}^\mu_{\htw} \equiv \frac{3}{2}S^{\mu}=0,\label{eq:MAGswelldDAFFEB}\\
        2\Check{E}_{\mu\nu} &\equiv& \Tilde{G}_{\mu\nu}-\frac{1}{2}g_{\mu\nu}\left(\hat{R}_T-2\Lambda + \frac32 S^2-\frac{1}{2}F_{(C)}^2\right)-\frac{3}{4}\mathring{\pi}{}_{\mu\alpha\beta}\mathring{\pi}{}_{\nu}{}^{\alpha\beta}+\nonumber\\
        &&+\mathring{q}{}_{\alpha\beta\mu}\mathring{q}{}^{\alpha\beta}{}_\nu+\mathring{t}{}_{\alpha\beta\mu}\mathring{t}{}^{\alpha\beta}{}_\nu+\frac12\left(\mathring{q}{}_{\mu\alpha\beta}\mathring{q}{}_{\nu}{}^{\alpha\beta}+\mathring{t}{}_{\mu\alpha\beta}\mathring{t}{}_{\nu}{}^{\alpha\beta}\right)+\nonumber\\
        &&+\mathring{t}{}_{\alpha\beta(\mu}\mathring{q}{}_{\nu)}{}^{\alpha\beta}-\mathring{q}{}^{\alpha\beta}{}_{(\mu}\mathring{t}{}_{\nu)\alpha\beta}-\mathring{t}{}_{\beta\alpha(\mu} \mathring{q}{}^{\alpha\beta}{}_{\nu)}+\frac{3}{2}S_\mu S_\nu- {F}^{(C)}_{\mu}{}^\alpha {F}^{(C)}_{\nu\alpha}.\label{eq:MAGswelldDAFFEmet}
    \end{eqnarray}
\end{subequations}
The first three of them imply $\mathring{O}{}_{\lambda\mu\nu}^{I}=0$. Then, equation $\mathcal{B}^\mu_{\htw}=0$ suggests that $S_\mu=0$ which yields $H_{\lambda\mu\nu}=0$ via~\eqref{def:PseudoVector}. Substituting these results back into \eqref{eq:MAGswelldDAFFE}, we obtain the effective set
\begin{subequations}\label{eq:MAGswellEffective}
\begin{eqnarray}
        \partial_\nu\left(\sqrt{-\mathsf g}{F}_{(C)}^{\nu\mu}\right)&=&0=\partial_\nu\left(\sqrt{-\mathsf g}\ast{F}_{(C)}^{\nu\mu}\right),\label{eq:MAGswellEqs}\\ 
        \tilde{G}_{\mu\nu}+\Lambda g_{\mu\nu}&=&{F}^{(C)}_{\mu}{}^\alpha {F}^{(C)}_{\nu\alpha}-\frac{1}{4}F_{(C)}^2g_{\mu\nu},\label{eq:MAGswellEffectiveFE}
\end{eqnarray}
\end{subequations}
where we took the liberty to also include the Bianchi identity for $F^{(C)}_{\mu\nu}$ with 
\begin{equation}\label{def:DualFC}
    \ast {F}_{(C)}^{\mu\nu}=\frac{1}{2}\Tilde{\epsilon}{}^{\mu\nu\rho\lambda}{F}^{(C)}_{\rho\lambda}.
\end{equation}
As differential equations, these exactly correspond to the Einstein-Maxwell system. Let us now study the solution in the different frameworks. 

\paragraph{Solution in the d-AF$^{\circ}$.} As already mentioned, we have $\mathring{O}^I_{\lambda\mu\nu}=0=S_\mu$ and  eq.~\eqref{eq:MAGswellEqs}, which only determines the combination $C_\mu$. If $\langle{C}_\mu\rangle+\partial_\mu\phi$ is the value of $C_\mu \equiv \alpha_A B^{A}_\mu$ satisfying~\eqref{eq:MAGswellEqs}, then $B^A$ acquires the value $\ev{B^A}$ with
\begin{equation}
    \ev{B^{\ho}_\mu} = \frac{\langle{C}_\mu\rangle + \partial_\mu\phi - \alpha_{\hz}\langle{B}^{\hz}_\mu\rangle}{\alpha_{\ho}}.
\end{equation}
Clearly, the values $\ev{B^A}+b^A$, where the $b^A$'s obey eq.~\eqref{eq:DAFgaugeMAGswell}, are also acceptable. A good strategy to capture the solution in all possible gauges is to set
\begin{equation}
    b^{\hz}=-\langle B^{\hz}\rangle + \tilde\alpha \langle C\rangle,\qquad \theta=-\phi,
\end{equation}
where $\tilde\alpha$ is a real number, obtaining
\begin{equation}\label{eq:dDAFsolutionMAGswell}
    B^{\hz} = \tilde\alpha \langle C\rangle,\qquad B^{\ho} = \frac{1-\tilde\alpha\alpha_{\hz}}{\alpha_{\ho}}\langle C\rangle.
\end{equation}
Therefore, 
\begin{equation}
    \{B^{\hz},B^{\ho}\} = \begin{cases}
        \{ 0, \ev{C}/\alpha_{\ho}\} & \tilde\alpha=0\\
        \{ \ev{C}/\alpha_{\hz},0\} & \tilde\alpha = 1/\alpha_{\hz}\\
        \{ \tilde\alpha\ev{C}, (1-\tilde\alpha\alpha_{\hz})\ev{C}/\alpha_{\ho}\} & \tilde\alpha\neq 0,1/\alpha_{\hz}.
    \end{cases}
\end{equation}

\paragraph{Solution in the AF.} The next step is to translate the solution into the language of the AF. The field equations tell us that $\mathring{O}^N_{\lambda\mu\nu}=0$, and that eq.~\eqref{eq:MAGswellEqs} must hold true. Again, if $\langle{C}_\mu\rangle+\partial_\mu\phi$ is the value of
\begin{equation}
    C_\mu \equiv \frac{\alpha_{\ho}^{}}{6 \sqrt{3}} \rho_{\mu } + \frac{2\alpha_{\hz}^{}}{3}  T_{\mu } + \frac{3 \alpha_{\hz}^{} + \sqrt{3} \alpha_{\ho}^{}}{9} u_{\mu }
\end{equation}
satisfying~\eqref{eq:MAGswellEqs}, then $A^I$ acquires the value $\ev{A^I}$ with
\begin{equation}
    \ev{u_\mu} = \frac{18\left(\langle C_\mu \rangle +\partial_\mu\phi\right)- \sqrt{3}\alpha_{\ho}\langle \rho_\mu\rangle - 12\alpha_{\hz}\langle T_\mu\rangle}{2(3\alpha_{\hz}+\alpha_{\ho}\sqrt{3})}.
\end{equation}
Clearly, the values $\ev{A^I}+b^{I-1}$, where the $b^{I-1}$'s obey eq.~\eqref{eq:AFgaugeMAGswell}, are as good. Setting
\begin{eqnarray}
    b^{\hz} = \alpha \ev{C} - \ev{T},\qquad b^{\ho} = \beta \ev{C} - \ev{\rho},\qquad \theta=-\phi,
\end{eqnarray}
where $\alpha,\beta$ are real numbers, we get 
\begin{equation}\label{eq:AFsolutionMAGswell}
    T =\alpha\langle C\rangle,\qquad \rho=\beta \langle C\rangle,\qquad u = \frac{18-12\alpha\alpha_{\hz}-\beta\alpha_{\ho}\sqrt{3}}{2(3\alpha_{\hz}+\alpha_{\ho}\sqrt{3})}\langle C\rangle.
\end{equation}
We can further identify
\begin{equation}
    \Tilde{\alpha} = \frac{18 + \sqrt{3} (4 \alpha -  \beta) \alpha_{\ho}^{}}{6 (3 \alpha_{\hz}^{} + \alpha_{\ho}^{} \sqrt{3})},
\end{equation}
and we collect the various cases in table~\ref{tab:AFtableMAGswell}.

\begin{table}[tbp]
\centering
\begin{tabular}{|ll|l|lll|}
\hline
$\alpha$&$\beta$&$\Tilde{\alpha}$&$T_\mu$&$\rho_\mu$&$u_\mu$\\
\hline 
0 & 0 & $\frac{3}{3\alpha_{\hz}+\alpha_{\ho}\sqrt{3}}$ & 0 & 0 & $\checkmark$\\
0 & $\frac{6\sqrt{3}}{\alpha_{\ho}}$ & 0 & 0 &$\checkmark$ &0\\
$\frac{3}{2\alpha_{\hz}}$ & 0 & $\frac{1}{\alpha_{\hz}}$ & $\checkmark$ & 0 & 0\\
$\neq 0,\frac{3}{2\alpha_{\hz}}$ & $\frac{2\sqrt{3}(3-2\alpha\alpha_{\hz})}{\alpha_{\ho}}$& $\frac{2\alpha}{3}$ &$\checkmark$&$\checkmark$&0\\
$\neq 0,\frac{3}{2\alpha_{\hz}}$ & 0 & $\frac{9 + 2 \sqrt{3} \alpha \alpha_{\ho}^{}}{9 \alpha_{\hz}^{} + 3  \alpha_{\ho}^{} \sqrt{3}}$ &$\checkmark$ & 0 &$\checkmark$\\
0&$\neq 0,\frac{6\sqrt{3}}{\alpha_{\ho}}$ & $\frac{18 -  \sqrt{3} \beta \alpha_{\ho}^{}}{6(3 \alpha_{\hz}^{} +  \alpha_{\ho}^{}\sqrt{3})}$ & 0 &$\checkmark$&$\checkmark$\\
$\neq 0$ & $\neq 0,\frac{2\sqrt{3}(3-2\alpha\alpha_{\hz})}{\alpha_{\ho}}$ & $\frac{18 + \sqrt{3} (4 \alpha -  \beta) \alpha_{\ho}^{}}{6 (3 \alpha_{\hz}^{} + \alpha_{\ho}^{} \sqrt{3})}$ & $\checkmark$ & $\checkmark$ & $\checkmark$\\
\hline
\end{tabular}
\caption{\label{tab:AFtableMAGswell} Values of $\Tilde{\alpha}$ and the AF field variables $A^I$ for different numbers $\alpha,\beta$. A checkmark $\checkmark$ indicates that the ticked field is proportional to $\langle C_\mu\rangle$.}
\end{table}

\paragraph{Solution in the FF.} The final step is to translate the solution into the language of the familiar Palatini formalism, i.e., to present an affine connection which solves the connection field equations. This reads
\begin{equation}
    \Gamma^\lambda{}_{\mu\nu}=\Tilde{\Gamma}^\lambda{}_{\mu\nu}+\langle V^\lambda\rangle g_{\mu\nu} - \delta^\lambda_\nu \frac{\langle C_\mu\rangle +\partial_\mu\phi -\left(\alpha_{\hz}+\alpha_{\ho}\sqrt{3}\right) \langle V_\mu\rangle}{\alpha_{\hz}-\alpha_{\ho}\sqrt{3}} + \delta^\lambda_\mu\langle U_\nu\rangle,
\end{equation}
where $\langle C_\mu\rangle +\partial_\mu\phi$ is the value of
\begin{equation}
    C_\mu\equiv \frac{3 \alpha_{\hz}^{} + 2 \sqrt{3} \alpha_{\ho}^{}}{9}  \check{Q}_{\mu } - \frac{6 \alpha_{\hz}^{} +  \sqrt{3} \alpha_{\ho}^{}}{18}  Q_{\mu } + \frac{2\alpha_{\hz}^{}}{3}  T_{\mu }
\end{equation}
satisfying~\eqref{eq:MAGswellEqs}. Again, due to the freedom to shift our connection as in~\eqref{eq:FFgaugeMAGswellShift}, and setting
\begin{subequations}
    \begin{eqnarray}
        a&=&\frac{24 -  (4 \alpha -  \beta) (\alpha_{\hz}^{} -   \alpha_{\ho} \sqrt{3}) }{12 (3 \alpha_{\hz}^{} +  \alpha_{\ho}\sqrt{3})}\langle C \rangle -  \langle V\rangle,\qquad \theta=-\phi,\\
        c &=& \frac{(8 \alpha + \beta) \alpha_{\hz}^2 - 4 \alpha_{\hz} (3 + 2  \alpha \alpha_{\ho}\sqrt{3}) - 3 \alpha_{\ho} (  \beta \alpha_{\ho}-4 \sqrt{3})}{12 (3 \alpha_{\hz}^2 - 2  \alpha_{\hz} \alpha_{\ho}\sqrt{3} - 3 \alpha_{\ho}^2)}\langle C\rangle -  \langle U\rangle,
    \end{eqnarray}
\end{subequations}
we reach a connection with torsion and non-metricity 
\begin{subequations}\label{eq:MAGswellSolFF}
    \begin{eqnarray}
        T^\lambda{}_{\mu\nu}&=&\frac{2\alpha}{3}\delta^\lambda_{[\nu}\langle C_{\mu]}\rangle,\\
        Q_{\lambda\mu\nu}&=&\frac{\beta}{6}g_{(\lambda\mu}\langle C_{\nu)}\rangle+\frac{18-12\alpha \alpha_{\hz}-\beta \alpha_{\ho}\sqrt{3}}{9(3\alpha_{\hz}+\alpha_{\ho}\sqrt{3})}\left( g_{\lambda(\mu}\langle C_{\nu)}\rangle - g_{\mu\nu}\langle C_\lambda\rangle\right).
    \end{eqnarray}
\end{subequations}
One can immediately verify that if we decompose the latter under the Lorentz group, we will exactly find~\eqref{eq:AFsolutionMAGswell} as the only excited irreducible modes. Therefore, one can again refer to table~\ref{tab:AFtableMAGswell} for the various cases.

An interesting remark is in order. The Lagrangian does undeniably propagate the massless combination $C_\mu$, a spin-1 geometric ``boson''. This means that part of the post-Riemannian structure gets (self-)excited but it turns out to be impossible to make a gauge-independent statement about specifically which part that is. For example, what appears to be an excitation of only torsional DoF in one gauge, shows up as an excitation of only non-metricity DoF in another. Hence, propagation of the MAGswell field is tantamount to a self-excitation of the connection background with different parts of the latter being excited in the different gauges.

Do also note that the action~\eqref{eq:MAGswellAction} can be thought of as the massless limit of a massive theory which has an action like~\eqref{eq:MAGswellAction}, but with $\mathcal{L}_{ct}$ replaced by 
\begin{equation}
    \mathcal{L}_{mass} = -\frac12\left[ (\mu^2\alpha_{\hz}^2-3)B^{\hz}_\mu B^{\hz}_\nu+2\mu^2\alpha_{\hz}\alpha_{\ho}B^{\hz}_\mu B^{\ho}_\nu+(\mu^2\alpha_{\ho}^2+3)B^{\ho}_\mu B^{\ho}_\nu \right]g^{\mu\nu},
\end{equation}
such that, up to surface terms, 
\begin{equation}\label{eq:MassiveVectorAction}
    I_{C}=\frac12 \int \sqrt{-\mathsf{g}}d^4x\left( \Tilde{R} -2\Lambda + \hat{R}_T + 3 S^2 -\mu^2C^2-\frac{1}{2}F_{(C)}^2\right),
\end{equation}
always in the d-AF$^{\circ}$. Obviously, $\mathcal{L}_{ct}=\mathcal{L}_{mass}(\mu=0)$. The last two terms in~\eqref{eq:MassiveVectorAction} imply that the combination $C_\mu$ behaves as a Proca field with mass $\mu$. Since the HP action already introduces a mass scale proportional to the Planck mass,\footnote{Remember that we have set the reduced Planck mass to unity.} naturalness criteria suggest that we take $\mu$ to be of the same order (the composite field $C_\mu$ is part of the space-time geometry, not some external field). The field equations in the d-AF$^{\circ}$ are~\eqref{eq:MAGswelldDAFFE} except that $\Tilde{\nabla}_\nu F_{(C)}^{\nu\mu}=0$ is replaced by the Proca equation 
\begin{equation}
    \Tilde{\nabla}_\nu F_{(C)}^{\nu\mu}-\mu^2 C^\mu=0.
\end{equation}
Observe that the massive action and the field equations following from it, do still possess a symmetry under~\eqref{eq:DAFgaugeMAGswellShift} if $\theta^A=0$. This means that the propagated combination $C_\mu$ is a \emph{massive} vector-boson, and the geometric interpretation of this propagation again falls into the previous scheme, viz, it is subject to the choice of gauge.  

Finally, we remark that the MAGswell field is of course by itself not a solid and unique concept. Nevertheless, let us justify why we think that $C_\mu$ is indeed the most general candidate to describe it. First of all, playing the devil's advocate, one could argue that there are more general projective-invariant combinations to take as our $B^A$ fields; we have already shown this in section~\ref{sec:HilbertNew}. Indeed, there is simply no physical argument favoring~\eqref{def:DAFBdoublet} over~\eqref{def:DAFBdoubletGeneral}. Sure, the diagonal form of the mass-squared matrix and the emergence of $SO$ symmetries under transformations in field space are nice features, but they are far from being necessary restrictions. Actually, these features are completely absent here because (i) we have removed the mass terms, and (ii) there are no $SO$ transformations being a symmetry of~\eqref{eq:MAGswellAction}. However, the real question is if we would gain more insight by considering a more complicated change of field variables. The answer is no, for $C_\mu$ would again be a linear combination of the $A^{I}$'s but with different coefficients. Assuming that we would also properly modify $\mathcal{L}_{ct}$ as to remove algebraic instances of the new $B^A$'s, it is evident that we would not get qualitatively different results. 

Second, one could argue that projective invariance of the action is definitely not mandatory. One could instead add a kinetic term for any of the vector variables in the AF and remove all algebraic instances of this field from the action by introducing a proper counter-term Lagrangian. This theory, which would no longer be invariant under~\eqref{eq:AFAtfs} (the AF analogue of what is a projective transformation in the FF), would then propagate the graviton and the specific mode. Fortunately, one can easily prove that the solutions in all these different cases would correspond to the one and only solution in the theory with action~\eqref{eq:MAGswellAction} in different gauges.    

\subsection{Non-linear interacting MAG theory, black holes and solitons}

To showcase the usefulness of this new approach to MAG for obtaining exact solutions with torsion and non-metricity, let us propose a four-dimensional interacting action with non-linear dynamics for the MAGswell field $C_\mu$ and the pseudo-vector $S_\mu$, namely
\begin{equation}\label{eq:NLaction}
    I_{NL} = \int \sqrt{-\mathsf{g}}d^4x\left[ \frac12(R - 2\Lambda - \hat{\mathcal{R}}_V) - \gamma_1 F_{(C)}^2 -\gamma_2 F_{(S)}^2 - \gamma \mathcal{L}_{int}\right],
\end{equation}
where 
\begin{equation}
    \mathcal{L}_{int} = \delta^{\mu_1...\mu_4}_{\nu_1...\nu_4} F^{(C)}_{\mu_1\mu_2}F^{(S)}_{\mu_3\mu_4}F_{(C)}^{\nu_1\nu_2}F_{(S)}^{\nu_3\nu_4},
\end{equation}
$\gamma$ is a positive coupling constant of mass-dimension $-4$, and $\gamma_1,\gamma_2$ are positive coupling constants of mass-dimension 0. This is the form of the action in the d-AF$^{\circ}$. The field strengths can be customarily written using only the partial derivative, i.e., $F_{(C)}$ as previously defined and $F^{(S)}_{\mu\nu}=\partial_{[\mu}S_{\nu]}$ with $S_\mu$ given in eq.~\eqref{def:PseudoVector}. The term $R$ stands for $\Tilde{R}+\hat{R}_T+\hat{\mathcal{R}}_V+\text{t.d.}$, with the constituents defined in eqs.~\eqref{eq:DAFscalConstituents}. One can always express the action in the FF (or the AF) by recalling eqs.~\eqref{eq:DAFtoAFtoFF}. This interacting Lagrangian was proposed in~\cite{Cisterna:2020rkc} for two distinct potentials in a Riemannian setup. Here, we endow these potentials with a special geometric origin and cast the whole thing as a MAG theory.

In the d-AF$^{\circ}$, the field equations read
\begin{subequations}\label{eq:NLdDAFFE}
    \begin{eqnarray}
        \mathring{\mathcal{O}}{}^{\lambda\mu\nu}_{\ho}&\equiv&\frac{1}{2}\left( \mathring{t}{}^{\lambda\mu\nu}-\mathring{q}{}^{[\mu\nu]\lambda}\right)=0,\qquad
        \mathring{\mathcal{O}}{}^{\lambda\mu\nu}_{\htw}\equiv-\frac{1}{4} \mathring{\pi}{}^{\lambda\mu\nu}=0,\nonumber\\
        \mathring{\mathcal O}{}^{\lambda\mu\nu}_{\hth}&\equiv&\mathring{\mathcal{O}}{}^{(\mu\nu)\lambda}_{\ho}+\frac{1}{8}\mathring{q}^{\lambda\mu\nu}=0,\\
        \frac{\mathcal{B}^\mu_{\hz}}{4\alpha_{\hz}}&\equiv&
    \frac{\mathcal{B}^\mu_{\ho}}{4\alpha_{\ho}}\equiv\gamma_1\Tilde{\nabla}_\nu{F}_{(C)}^{\nu\mu} - \gamma \delta^{\mu\nu\rho\sigma}_{\alpha\beta\gamma\delta}F^{(S)}_{\rho\sigma}\Tilde{\nabla}_{\nu}\left( F_{(C)}^{\alpha\beta}F_{(S)}^{\gamma\delta}\right)=0,\label{eq:NLdDAFFEB12}\\
    \frac{\mathcal{B}^\mu_{\htw}}{4}&\equiv& \gamma_2\Tilde{\nabla}_\nu F_{(S)}^{\nu\mu} + \gamma \delta^{\nu\rho\sigma\mu}_{\alpha\beta\gamma\delta} F^{(C)}_{\nu\rho}\Tilde{\nabla}_{\sigma}\left(F_{(C)}^{\alpha\beta}F_{(S)}^{\gamma\delta}\right)=0,\label{eq:NLdDAFFEB3}\\
    2\Check{E}_{\mu\nu} &\equiv& \Tilde{G}_{\mu\nu}-\frac{1}{2}g_{\mu\nu}\left[\hat{R}_T-2\Lambda -2\gamma_1 F_{(C)}^2 -2\gamma_2 F_{(S)}^2+ 2 \gamma \mathcal{L}_{int}\right]+\nonumber\\
        &&+\mathring{q}{}_{\alpha\beta\mu}\mathring{q}{}^{\alpha\beta}{}_\nu+\mathring{t}{}_{\alpha\beta\mu}\mathring{t}{}^{\alpha\beta}{}_\nu+\frac12\left(\mathring{q}{}_{\mu\alpha\beta}\mathring{q}{}_{\nu}{}^{\alpha\beta}+\mathring{t}{}_{\mu\alpha\beta}\mathring{t}{}_{\nu}{}^{\alpha\beta}\right)+\nonumber\\
        &&+\mathring{t}{}_{\alpha\beta(\mu}\mathring{q}{}_{\nu)}{}^{\alpha\beta}-\mathring{q}{}^{\alpha\beta}{}_{(\mu}\mathring{t}{}_{\nu)\alpha\beta}-\mathring{t}{}_{\beta\alpha(\mu} \mathring{q}{}^{\alpha\beta}{}_{\nu)}-\frac{3}{4}\mathring{\pi}{}_{\mu\alpha\beta}\mathring{\pi}{}_{\nu}{}^{\alpha\beta}-\nonumber\\
        &&-4\gamma_1{F}^{(C)}_{\mu}{}^\alpha {F}^{(C)}_{\nu\alpha}- 4\gamma_2{F}^{(S)}_{\mu}{}^\alpha {F}^{(S)}_{\nu\alpha}=0.\label{eq:NLdDAFFEmet}
    \end{eqnarray}
\end{subequations}
To get the above expressions, we also used the Bianchi identities $dF_{(C)}=0=dF_{(S)}$ and the dimension-dependent identity 
\begin{equation}
    \delta^{\mu_1 ... \mu_4}_{\nu_1...\nu_4}F^{(C)}_{[\mu_1\mu_2}F^{(S)}_{\mu_3\mu_4}F_{(C)}^{\nu_1\nu_2} F_{(S)}^{\nu_3\nu_4}g_{\mu]\nu}=0.
\end{equation}

The action and the field equations are invariant under the transformation~\eqref{eq:DAFgaugeMAGswellShift} with the $b^A$'s constrained via~\eqref{eq:DAFgaugeMAGswell}. They are also invariant under a shift of $S_\mu$ by a locally exact co-vector, i.e., $S'_\mu = S_\mu + \partial_\mu\phi$, which in the AF amounts to
\begin{equation}
    H'_{\lambda\mu\nu}= H_{\lambda\mu\nu} - \Tilde{\epsilon}_{\lambda\mu\nu\alpha}\partial^\alpha\phi,
\end{equation}
whereas it corresponds to the connection transformation
\begin{equation}
    \Gamma'^\lambda{}_{\mu\nu}=\Gamma^\lambda{}_{\mu\nu}+\frac12 \Tilde{\epsilon}^\lambda{}_{\mu\nu\alpha}\partial^\alpha\phi,
\end{equation}
in the FF. Following the steps laid down in the previous section, we expect a connection solution with torsion and non-metricity
\begin{subequations}
    \begin{eqnarray}
        T^\lambda{}_{\mu\nu}&=&\frac{2\alpha}{3}\delta^\lambda_{[\nu}\langle C_{\mu]}\rangle + \ev{S_\alpha}\Tilde{\epsilon}^{\alpha\lambda}{}_{\mu\nu},\\
        Q_{\lambda\mu\nu}&=&\frac{\beta}{6}g_{(\lambda\mu}\langle C_{\nu)}\rangle+\frac{18-12\alpha \alpha_{\hz}-\beta \alpha_{\ho}\sqrt{3}}{9(3\alpha_{\hz}+\alpha_{\ho}\sqrt{3})}\left( g_{\lambda(\mu}\langle C_{\nu)}\rangle - g_{\mu\nu}\langle C_\lambda\rangle\right),
    \end{eqnarray}
\end{subequations}
respectively, where $\ev{C}$ and $\ev{S}$ satisfy the non-linear differential equations~\eqref{eq:NLdDAFFEB12} and~\eqref{eq:NLdDAFFEB3}.  We remind the reader that our gauge freedom is fully exhausted once we fix values for $\alpha,\beta$ (see table~\ref{tab:AFtableMAGswell}). 

Now, let us consider the static spherically-symmetric metric ansatz 
\begin{equation}
    ds^2=-f(r)dt^2 + \frac{dr^2}{f(r)}+r^2d \Sigma^2_2,
\end{equation}
where $d \Sigma^2_2=d\chi^2 + \sin^2\chi dy^2$ gives the line element of a two-dimensional spherical section with $\chi,y$ compact. We also make the following ans\"atze,  
\begin{equation}\label{eq:NLansatze}
    C_\mu = c(r)\delta^0_\mu,\qquad S_\mu = p \cos\chi \delta^{3}_\mu,
\end{equation}
which result in
\begin{equation}
    F^{(C)}_{\mu\nu} = c'\delta^{10}_{\mu\nu},\qquad F^{(S)}_{\mu\nu} = p\sin\chi \delta^{32}_{\mu\nu}.
\end{equation}
A prime denotes differentiation with respect to $r$. 

Given the above, eq.~\eqref{eq:NLdDAFFEB3} is identically satisfied, while eq.~\eqref{eq:NLdDAFFEB12} gives 
\begin{equation}
    r(8\gamma p^2 + \gamma_1 r^4)c'' + 2(\gamma_1 r^4-8\gamma p^2)c'=0.
\end{equation}
This yields the first integral
\begin{equation}
    c' = -\frac{q r^2}{\gamma_1 r^4+ 8 \gamma p^2},
\end{equation}
where $q$ is an integration constant. Integrating once more, we get
\begin{equation}\label{eq:cexpression}
    c = c_0 +\frac{q}{\gamma_1 r}{_2F_1}\left[ \frac{1}{4},1,\frac54;-\frac{8\gamma p^2}{\gamma_1 r^4}\right],
\end{equation}
where ${_2F_1}$ is the Gaussian hypergeometric function~\cite{abramowitz+stegun}, and $c_0$ is another constant of integration. Therefore our connection solution is such that its torsion and non-metricity read 
\begin{subequations}\label{eq:NLFFsol}
    \begin{eqnarray}
        T^\lambda{}_{\mu\nu} &=& -\frac{\alpha}{3}\delta^{\lambda 0}_{\mu\nu} c + p\cos\chi\Tilde{\epsilon}^{3\lambda}{}_{\mu\nu},\\
        Q_{\lambda\mu\nu} &=& c\left[\frac{\beta}{6}g_{(\lambda\mu}\delta^0_{\nu)}+\frac{18-12\alpha \alpha_{\hz}-\beta \alpha_{\ho}\sqrt{3}}{9(3\alpha_{\hz}+\alpha_{\ho}\sqrt{3})}\left( g_{\lambda(\mu}\delta^0_{\nu)} - g_{\mu\nu}\delta^0_\lambda\right)\right],
    \end{eqnarray}
\end{subequations}
with $c$ given in~\eqref{eq:cexpression}. 

Plugging this into~\eqref{eq:NLdDAFFEmet}, we find that 
\begin{eqnarray}\label{eq:ttcomponent}
    -\frac{2r}{f}\Check{E}_{00} = f' + \frac{f}{r} -\frac{k}{r} + \Lambda r +\frac{2\gamma_2 p^2}{r^3} + \frac{2q^2 r}{\gamma_1 r^4 + 8\gamma p^2}=0.
\end{eqnarray}
Since 
\begin{equation}
    \Check{E}_{11}=-f^{-2}\check{E}_{00},\qquad \Check{E}_{33}=\sin^2\chi\Check{E}_{22},
\end{equation}
and 
\begin{equation}
    \Check{E}_{22}=\frac{r^2}{2f}\Check{E}_{00}-\left( \frac{r^3}{2f} \Check{E}_{00}  \right)',
\end{equation}
we only have to find the solution to eq.~\eqref{eq:ttcomponent}, which reads
\begin{equation}\label{eq:MetricBack}
    f=1-\frac{2M}{r}-\frac{\Lambda r^2}{3} + \frac{2\gamma_2 p^2}{r^2} + \frac{2 q^2}{\gamma_1 r^2} {_2F_1}\left[\frac14,1,\frac54;-\frac{8\gamma p^2}{\gamma_1 r^4} \right].
\end{equation}
The symbol $M$ stands for yet another integration constant, this time associated with the mass. The very interesting metric background~\eqref{eq:MetricBack} has been extensively studied in~\cite{Liu:2019rib,Cisterna:2020rkc}, and there is no need to discuss it here in depth. In our case, nevertheless, the corrections to the Schwarzschild-(A)dS metric is due to a richer space-time geometry and not due to the introduction of additional fields (like a Maxwell field). In this sense, this is a novel result.

Some comments are in order. Observe that by setting $p=0$, the background~\eqref{eq:MetricBack} assumes the form
\begin{equation}
    f = 1 -\frac{2M}{r} + \frac{8q^2}{r^2} - \frac{\Lambda r^2}{3}, 
\end{equation}
and, up to choice of the integration constant $q$, it is indeed the metric solution in the MAGswell theory with action~\eqref{eq:MAGswellAction} if we make the ansatz~\eqref{eq:NLansatze} for $C_\mu$. Moreover, the torsion and non-metricity of the connection solution, eqs.~\eqref{eq:NLFFsol}, acquire the form~\eqref{eq:MAGswellSolFF}, ergo, we recover the full solution in the MAGswell model, as a special case. Another interesting setup is to consider the action~\eqref{eq:NLaction} with $\Lambda=0=\gamma_1$. In this case, the connection solution will have torsion and non-metricity~\eqref{eq:NLFFsol} with 
\begin{equation}
    c= c_0 + q r^3,
\end{equation}
whereas the metric function $f$ will be 
\begin{equation}
    f = 1-\frac{2M}{r}+\frac{2\gamma_2 p^2}{r^2}-\frac{\Lambda_{eff}r^2}{3},
\end{equation}
where $\Lambda_{eff}>0$ stands for the effective cosmological constant
\begin{equation}
    \Lambda_{eff} = 16\gamma p^2 q^2.
\end{equation}

Finally, non-singular solutions were reported in~\cite{Cisterna:2020rkc} for a specific choice of the mass parameter $M$ in a strongly-coupled regime. The need to go to such a regime will not be necessary here; we will just set $\gamma_2=0$ and choose our mass parameter as  
\begin{equation}
    M=M_{*}:=\frac{\pi q^2}{4(2\gamma p^2 \gamma_1^3)^{1/4}}.
\end{equation}
Then, eq.~\eqref{eq:MetricBack} assumes the expression
\begin{equation}\label{eq:RegularMetricBack}
    f = 1-\frac{2M_*}{r}-\frac{\Lambda r^2}{3} + \frac{2q^2}{\gamma_1 r^2}{_2F_1\left[\frac14,1,\frac54; -\frac{8\gamma p^2}{\gamma_1 r^4} \right]},
\end{equation}
and admits the near-origin expansion
\begin{equation}
    f\underset{r\to 0}{=} 1 - \left( \frac{q^2}{12\gamma p^2}+\frac{\Lambda}{3}\right)r^2 + \mathcal{O}(r^3).
\end{equation}
If $\Lambda\geq 0$ or $-q^2/(4\gamma p^2)<\Lambda<0$, the presence of a de Sitter core with radius
\begin{equation}
    l_{dS} = \frac{2p\sqrt{3\gamma}}{\sqrt{q^2+4\gamma p^2\Lambda}},
\end{equation}
is manifest, ensuring regularity of Riemann-curvature invariants at the origin and completeness in the geodesic sense~\cite{1966JETP...22..241S}. A further study of the causal structure of the solution reveals~\cite{Cisterna:2020rkc} that, for certain values (or ranges thereof) of the coupling/integration constants, eq.~\eqref{eq:RegularMetricBack} describes either a gravitational soliton (horizon-free solution with regular origin), or just a standard black hole solution with an extremal limit. To the best of our knowledge, regular black hole solutions with torsion and non-metricity have not been yet reported in the MAG literature. 

Since the actual novelty in the full solution is the existence of a non-trivial connection background, we find it worth to include a few lines about the behavior of the latter in various limits. First, let us write torsion and non-metricity in a coordinate-free manner by introducing a vierbein field $e_a^\mu$, with indices $a,b,... = (0),...,(3)$ and inverse $e^a_\mu$ satisfying the orthonormality relation $g_{\mu\nu}=\eta_{ab}e^a_\mu e^b_\nu$. In particular, let us choose it to be diagonal, viz., 
\begin{equation}
    e^a_\mu = \mathrm{diag}\left( \sqrt{f},\frac{1}{\sqrt{f}},r,r\sin\chi\right).
\end{equation}
Then, the only non-vanishing components of $T^a{}_{bc}=e^a_\lambda e_b^\mu e_c^{\nu}T^\lambda{}_{\mu\nu}$ are
\begin{eqnarray}
    T^{(0)}{}_{(1)(2)}&=&T^{(1)}{}_{(0)(2)}=-T^{(2)}{}_{(0)(1)}=\frac{p\cot\chi}{r},\qquad T^{(i)}{}_{(0)(i)}=\frac{\alpha c}{3\sqrt{f}},
\end{eqnarray}
where $i,j,...$ take values in $\{1,2,3\}$. It seems that the $(i)(0)(i)$ components will be singular at the horizon radius $r=r_+$. Fortunately, this can be remedied by fixing the integration constants $c_0$ in~\eqref{eq:cexpression} as
\begin{equation}
    c_0 = - \frac{q}{\gamma_1 r_+}{_2F_1}\left[ \frac{1}{4},1,\frac54;-\frac{8\gamma p^2}{\gamma_1 r_+^4}\right],
\end{equation}
so that $c \sim (r-r_+)$ near the horizon surface. Moreover, all components of the torsion tensor exhibit a $~r^{-1}$ fall-off at asymptotic infinity. Next, we have a single pole at the origin $r=0$ due to the axial part. This pole persists even in the case of the regular metric solution~\eqref{eq:RegularMetricBack}. If we assume that a probe particle with micro-structure follows the auto-parallels, then it is a good question to ask whether this particle is going to ``feel'' the torsion singularity at the origin. Thankfully, the axial part of torsion drops out of the auto-parallel equation~\cite{Iosifidis:2019dua}, and thus, this singular behavior should not really be a cause for concern! Finally, all components of $Q_{abc}=e_a^\lambda e_b^{\mu} e_c^\nu Q_{\lambda\mu\nu}$ are proportional to $c/\sqrt{f}$. For $f$ as in~\eqref{eq:MetricBack}, this ratio vanishes at all previously discussed radii. On the other hand, in the case of the regular metric~\eqref{eq:RegularMetricBack}, it acquires a finite value at the origin. In the regular extremal case, it further is finite also at $r=r_+$.

\subsection{Cosmological constant powered by torsion}
It is an old fact that the minimal coupling of a 3-form field to Einstein gravity without a cosmological constant leads to Einstein's field equations with a cosmological constant purely derived from a gauge principle~\cite{Duff:1980qv}. Here, we shall disclose a MAG model with no cosmological constant which also leads to pure gravity with a cosmological constant, the latter now powered by axial torsion. 

Let us consider the projective-invariant action
\begin{equation}\label{eq:MAGcosmo}
    I_H = \frac{1}{2}\int \sqrt{-\mathsf{g}}d^4x \left( R + \frac{1}{4}H^2 - \frac{1}{24}F_{(H)}^2\right),
\end{equation}
where $F^{(H)}_{\lambda\mu\nu\rho}= 4 \partial_{[\lambda}H_{\mu\nu\rho]}$. The purpose of the second term in the above integral is to cancel out the mass term for $H_{\lambda\mu\nu}$ present in the AF expression of the Ricci scalar. This ensures that the action~\eqref{eq:MAGcosmo} is invariant under the translation 
\begin{equation}
    H'{}_{\lambda\mu\nu} = H_{\lambda\mu\nu} + \partial_{[\lambda}B_{\mu\nu]}
\end{equation}
which corresponds to the transformation
\begin{equation}
    \Gamma'{}^\lambda{}_{\mu\nu} = \Gamma^\lambda{}_{\mu\nu} - \frac12 g^{\lambda\rho}\partial_{[\rho}B_{\mu\nu]}
\end{equation}
of the affine connection in the FF, with $B_{\mu\nu}$ being an arbitrary 2-form field.

In the convenient AF$^{\circ}$, the field equations read
\begin{subequations}
    \begin{eqnarray}
        \mathring{\mathcal{O}}^{\lambda\mu\nu}{}_{\hz}&=&\frac{1}{6} \Tilde{\nabla}_\alpha {F}_{(H)}^{\alpha\lambda\mu\nu}=0,\\
        \mathring{\mathcal{O}}{}^{\lambda\mu\nu}_{\ho}&\equiv&\frac{1}{2}\left( \mathring{t}{}^{\lambda\mu\nu}-\mathring{q}{}^{[\mu\nu]\lambda}\right)=0,\qquad
        \mathring{\mathcal{O}}{}^{\lambda\mu\nu}_{\htw}\equiv-\frac{1}{4} \mathring{\pi}{}^{\lambda\mu\nu}=0,\nonumber\\
        \mathring{\mathcal O}{}^{\lambda\mu\nu}_{\hth}&\equiv&\mathring{\mathcal{O}}{}^{(\mu\nu)\lambda}_{\ho}+\frac{1}{8}\mathring{q}^{\lambda\mu\nu}=0,\\
        \mathcal{B}^{\mu}_{\hz}&\equiv& -\frac{3}{2}B^\mu_{\hz}=0, \qquad \mathcal{B}^{\mu}_{\ho}\equiv \frac{3}{2}B^\mu_{\ho}=0, \\
        2\hat{E}_{\mu\nu} &=& \Tilde{G}_{\mu\nu}-\frac{1}{2}g_{\mu\nu}\left(R_T+\mathcal{R}_V+ \frac{1}{4}H^2 - \frac{1}{24} F_{(H)}^2\right)-\frac{3}{4}\mathring{\pi}{}_{\mu\alpha\beta}\mathring{\pi}{}_{\nu}{}^{\alpha\beta}+\nonumber\\
        &&+\mathring{q}{}_{\alpha\beta\mu}\mathring{q}{}^{\alpha\beta}{}_\nu+\frac{1}{2}\mathring{q}{}_{\mu\alpha\beta}\mathring{q}{}_{\nu}{}^{\alpha\beta}+\mathring{t}{}_{\alpha\beta\mu}\mathring{t}{}^{\alpha\beta}{}_\nu+\frac{1}{2}\mathring{t}{}_{\mu\alpha\beta}\mathring{t}{}_{\nu}{}^{\alpha\beta}+\mathring{t}{}_{\alpha\beta(\mu}\mathring{q}{}_{\nu)}{}^{\alpha\beta}-\nonumber\\
        &&-\mathring{q}{}^{\alpha\beta}{}_{(\mu}\mathring{t}{}_{\nu)\alpha\beta}-\mathring{t}{}_{\beta\alpha(\mu} \mathring{q}{}^{\alpha\beta}{}_{\nu)}+\frac{3}{2}B_\mu^{A}B_\nu^{B}\eta_{AB}-\frac{1}{6}F^{(H)}_{\mu}{}^{\alpha\beta\gamma}F^{(H)}_{\nu\alpha\beta\gamma}=0
    \end{eqnarray}
\end{subequations}
where the expressions of the $B_\mu^{A}$'s in terms of the AF (or FF) variables are to be found in eqs.~\eqref{eq:DAFtoAFtoFF}, $R_T$ in~\eqref{eq:AFscalConstituents} (first one), and $\mathcal{R}_V$ in~\eqref{eq:DAFRV}. From the above, it is quite easy to conclude that $\mathring{O}^I=0=B^A$. Therefore, the field equations assume the effective form
\begin{subequations}
    \begin{eqnarray}
         \Tilde{\nabla}_\alpha {F}_{(H)}^{\alpha\lambda\mu\nu}&=&0,\label{eq:DAFeffectiveH}\\
         \Tilde{G}_{\mu\nu}&=&\frac{1}{48}\left( 8{F}^{(H)}_\mu{}^{\alpha\beta\gamma}{F}^{(H)}_{\nu\alpha\beta\gamma} - g_{\mu\nu}F_{(H)}^2\right).\label{eq:DAFeffectiveMet}
    \end{eqnarray}
\end{subequations}

To proceed, one must now recall that 
\begin{equation}
    F^{(H)}_{\lambda\rho\mu\nu}=\chi(x)\Tilde{\epsilon}_{\lambda\rho\mu\nu},
\end{equation}
since $F_{(H)}$ is a top-form in four dimensions. Clearly, equation~\eqref{eq:DAFeffectiveH} implies that $\chi$ is an integration constant, say equal to $\chi_0$. Consequently, we are left with 
\begin{equation}
    \Tilde{G}_{\mu\nu}+\frac{1}{2}\chi_0^2 g_{\mu\nu}=0,
\end{equation}
which will determine the metric, and we directly find that 
\begin{equation}
    \Tilde{R}_{\mu\nu} = \frac{\chi_0^2}{2}g_{\mu\nu},
\end{equation}
which is the familiar Riemannian Ricci-curvature condition for Einstein manifolds with positive constant curvature. 

As promised, we found a connection solution which features only axial torsion $H_{\lambda\mu\nu}$. Again, we stress out that this type of torsion has no effect on the auto-parallels, i.e., the latter continue to coincide with the geodesics. We also saw that our field equations do effectively become Einstein's field equations with a positive (effective) cosmological constant, $\Lambda_{eff}=\chi_0^2/2$, once we integrate out the connection. The cosmological solution in the absence of matter sources would then be a de Sitter universe with Hubble constant $H \propto |\chi_0|$ where the expansion is now driven by an actual integration constant, powered by torsion, instead of an a priori fixed value.  

\section{Summary and future prospects}
\label{sec:Conclusions}

We started from the observation that the affine connection is a single field encoding $n^3$-many off-shell DoF, arguing that, for certain purposes, it might be more efficient to distribute these degrees among more than one fields. We then proceeded with a convenient change of field variables $\{g,\Gamma\}\to \{ g,\mathring{O}^N,A^I\}$ going to a framework which we dubbed AF. Besides the metric, the new field variables are the irreducible pieces of the torsion and non-metricity tensors under the Lorentz group. They are thus automatically identified with the fundamental fields $\{g,\Gamma\}$ in the FF. 

We worked out in detail the relations between the functional derivatives in the two frameworks and concluded that, not surprisingly, the field equations in the AF imply and are implied by the field equations in the FF. Hence, the field equations in the AF constitute an equivalent system, and we have the freedom, at any stage, to switch between the different frameworks. To complete the mapping, we further disclosed a correspondence between linear connection transformations in the FF and translations in the AF while we also determined how the $\mathring{O}^{N}$'s and the $A^{I}$'s should transform under a local Weyl re-scaling of the metric.  

We then applied the AF to the Hilbert-Palatini action and showed its well-known equivalence to Einstein gravity (up to choice of a gauge) also in the new framework. Observing that a projective transformation of the connection corresponds to simultaneous translations of the $A^I$'s in the AF, we further argued that the projective symmetry manifests itself as a true gauge symmetry in the new framework, i.e., one of the components of the vector triplet $A_\mu$ is redundant. In particular, this means that any projective-invariant action admits a description in terms of a reduced set of variables $\{g,\mathring{O}^N,B^A\}$ where the $B^A$'s are in general identified with linear combinations of the $A^I$'s. This led us to develop a useful variant of the AF, which we dubbed diminished AF or AF$^{\circ}$ for short. 

We saw that there exists a particular choice of combinations $B^A$ which reveals an $SO(1,1)$ symmetry of the $n$-dimensional HP action under a group action on the components of the doublet $B_\mu$. In $n=4$, the field variables in the AF$^{\circ}$ can be re-organized. Using the fact that the dual of the 3-form torsion is a pseudo-vector, the quadruplet $\mathring{O}_{\lambda\mu\nu}$ is reduced to a triplet by handing its first component to the doublet $B_\mu$ which becomes a triplet. This is just a special four-dimensional variant of the AF$^{\circ}$, obtained via the change of variables $\{g,\mathring{O}^N,B^A\}\to \{ g, \mathring{O}^I, B^{\mathcal{A}}\}$, which we called d-AF$^{\circ}$ for the sake of clarity. As it turns out, the HP action proves to be an $SO(1,2)$-symmetric action in the d-AF$^{\circ}$ where the group action mixes the $B^{\mathcal{A}}$'s. Actions of the $SO(2)$ subgroup rotate the elements of a two-dimensional subspace with the discrete version for $\theta=\pi/2$ interpreted as a rotation of axial torsion to non-metricity, and vice versa. 

Observing that any MAG theory in these alternative frameworks can be handled as a Riemannian theory with additional fields, we argued that it is an efficient strategy to use solvable (and suitable) Riemannian theories as ``seeds'' for solvable MAG theories which propagate the connection in vacuum. As our first example, we drew inspiration from the elegant Einstein-Maxwell theory. We proposed a theory for what we called the MAGswell field, a composite field labeling a projective-invariant linear combination of torsion and non-metricity traces. The naive action should follow from the Maxwell action by replacing $\Tilde{R}$ with $R$ and the gauge field with the MAGswell field, the latter having nothing to do with a gauge connection. Doing so, one of course notices that what was a $U(1)$ of the second kind in the Riemannian case does not translate into a symmetry of the MAG theory under locally exact shifts of the MAGswell field. The reason is that the presence of the Ricci scalar makes the field massive. Thus, a counter-term Lagrangian was also included with the sole purpose of removing the mass terms for the constituents of the composite MAGswell field. 

We then discussed the symmetries of the MAGswell action in all frameworks. Exactly because the MAGswell field is a composite object, we showed that the action is symmetric under a 2-parameter transformation of the vector variables in the AF$^{\circ}$ (or the d-AF$^{\circ}$), which combines a transformation preserving the MAGswell field and one translating it by an exact vector. In the other frameworks, this symmetry shows up as a symmetry under a 3-parameter transformation, a fact attributed to the absorption of the projective-symmetry charge in the diminished AF. We derived the field equations in the d-AF$^{\circ}$ and presented the solution in all frameworks, finding a proper expression that captures its form in all gauges. Actually, the reader was provided with table~\ref{tab:AFtableMAGswell} which displays all cases possible, and which proves that the propagation of the MAGswell field, a gauge-independent fact, cannot be tied to a self-excitation of a uniquely determined part of the post-Riemannian structure in a gauge-independent fashion, i.e., different parts of the connection background get excited for different choices of gauge. 

After this instructive example, we proceeded with a more complicated theory, this time inspired by quasi-topological electromagnetism~\cite{Cisterna:2020rkc}. We proposed a Lagrangian with non-linear dynamics for the MAGswell field and the torsion pseudo-vector letting them interact with each other. After briefly discussing the symmetries and deriving the field equations, we adopted a static and spherically-symmetric metric ansatz, together with compatible ans\"atze for torsion and non-metricity, in an attempt to recover the black hole solution reported in~\cite{Cisterna:2020rkc,Liu:2019rib}. The full solution describes a black hole with a non-zero connection background sourcing the post-Schwarzschild contributions to the metric solution. Under a certain tuning of the integration constants, we also showed that this black hole exhibits a regular core and is thus complete in the geodesic sense. 

However, assuming that particles with micro-structure follow auto-parallels, we also had to analyze the behavior of the torsion and non-metricity of the solution at all radii of interest. Doing so, we had to fix yet another integration constant to avoid a singular behavior at the horizon radius, but we concluded that there is no remedy for a single pole at the origin due to axial torsion. This pole is inevitable even in the case of the regular black hole. Nevertheless, as we pointed out, the axial piece of torsion drops out from the auto-parallel equation meaning that the probe particle would never be affected by this torsion singularity. 

Finally, as our last example, and inspired by the derivation of a cosmological constant from a gauge principle~\cite{Duff:1980qv}, we put forth a simple MAG action for the 3-form torsion. After deriving the field equations we presented a connection solution featuring only axial torsion which powers a positive effective cosmological constant. The cosmological solution in this MAG theory --- in the absence of matter sources --- would be a de Sitter universe with the expansion driven by torsion. We remarked that the effective cosmological constant is an integration constant as opposed to a fixed-value $\Lambda$ introduced by hand in the action.

The main goal of this work was to communicate the idea that a smart change of field variables can be a really useful strategy when trying to find solvable MAG theories. Indeed, our proposal proves to be a fruitful one, for although we restricted ourselves to showing only three examples, these are suggestive of many more. Writing down simple field theories for the new field variables compared to considering combinations of curvature invariants to make the connection dynamical, is of course a far less general method, albeit a much more targeted and result-oriented one. In the future, we plan to give more examples and solutions which are not necessarily inspired by Riemannian theories. We find it interesting to study kinetic theories for the various tensor modes and also investigate if (and how) Riemannian theories with scalar fields can fit as an inspiration into this description of MAG.  

\appendix
\section{Irreducible decomposition of a rank-3 tensor}
The irreducible decomposition of a general rank-3 tensor $\Delta_{\lambda\mu\nu}$ under the Lorentz group reads 
\begin{eqnarray}\label{eq:DeltaIRD}
    \Delta_{\lambda\mu\nu} &=& \Delta_{[\lambda\mu\nu]}+\mathring{\Delta}{}_{(\lambda\mu\nu)}+\mathring{D}{}_{\lambda[\mu\nu]}+\mathring{D}{}_{\lambda(\mu\nu)}+\bar{\Delta}{}_{(\lambda\mu\nu)}+\bar{D}{}_{\lambda[\mu\nu]}+\bar{D}{}_{\lambda(\mu\nu)},
\end{eqnarray}
where
\begin{subequations}\label{eq:DeltaIRDconstituents}
    \begin{eqnarray}
        \mathring{D}{}_{\lambda\mu\nu}&=&\Delta_{\lambda\mu\nu}- \Delta_{(\lambda\mu\nu)}-\Delta_{[\lambda\mu\nu]}-\frac{1}{n-1}g_{\lambda[\mu}\left(\Delta^\alpha{}_{|\alpha|\nu]} - \Delta^\alpha{}_{\nu]\alpha}\right)-\nonumber\\
        &&-\frac{1}{3(n-1)} g_{\lambda(\mu}\left( \Delta^\alpha{}_{|\alpha|\nu)}+\Delta^\alpha{}_{\nu)\alpha}-2\Delta_{\nu)\alpha}{}^\alpha\right)  +\nonumber\\
        &&+\frac{1}{3(n-1)} g_{\mu\nu}\left( \Delta^\alpha{}_{\alpha\lambda}+\Delta^\alpha{}_{\lambda\alpha}-2\Delta_{\lambda\alpha}{}^\alpha\right),\\
        \mathring{\Delta}{}_{(\lambda\mu\nu)}&=& \Delta_{(\lambda\mu\nu)}-\frac{1}{D+2}g_{(\mu\nu}\left(\Delta_{\lambda)\alpha}{}^\alpha + \Delta^\alpha{}_{\lambda)\alpha}+\Delta^\alpha{}_{|\alpha|\lambda)}\right),\\
        \bar{\Delta}{}_{(\lambda\mu\nu)}&=&\frac{1}{n+2}g_{(\mu\nu}\left(\Delta_{\lambda)\alpha}{}^\alpha + \Delta^\alpha{}_{\lambda)\alpha}+\Delta^\alpha{}_{|\alpha|\lambda)}\right)=\Delta_{(\lambda\mu\nu)}-\mathring{\Delta}{}_{(\lambda\mu\nu)},\\
        \bar{D}_{\lambda\mu\nu}&=&\frac{1}{n-1}g_{\lambda[\mu}\left(\Delta^\alpha{}_{|\alpha|\nu]} - \Delta^\alpha{}_{\nu]\alpha}\right)-\frac{1}{3(n-1)} g_{\mu\nu}\left( \Delta^\alpha{}_{\alpha\lambda}+\Delta^\alpha{}_{\lambda\alpha}-2\Delta_{\lambda\alpha}{}^\alpha\right)+\nonumber\\
        &&+\frac{1}{3(n-1)} g_{\lambda(\mu}\left( \Delta^\alpha{}_{|\alpha|\nu)}+\Delta^\alpha{}_{\nu)\alpha}-2\Delta_{\nu)\alpha}{}^\alpha\right)=D_{\lambda\mu\nu} - \mathring{D}{}_{\lambda\mu\nu}.
    \end{eqnarray}
\end{subequations}

\section{Glossary}
\begin{table}[htp]
\centering
\begin{tabular}{|l|l|}
\hline
Indices&Values\\
\hline 
$\mu,\nu,...$ & 0,1,...,$n-1$\\
$i,j,...$ & 1,2,...,$n-1$\\
$a,b,...$ & (0),(1),...,($n-1$)\\
$M,N,...$ & $\hz,\ho,\htw,\hth$\\
$I,J,...$ & $\ho,\htw,\hth$\\
$A,B,...$ & $\hz,\ho$\\
$\mathcal{A},\mathcal{B},...$ & $\hz,\ho,\htw$\\
\hline
\end{tabular}
\caption{\label{tab:Indices} Indices used in this work and their values.}
\end{table}

\begin{table}[htp]
\centering
\begin{tabular}{|l|l|}
\hline
Acronym&Full name\\
\hline 
MAG & Metric-affine gravity\\
DoF & Degrees of freedom\\
FF & Fundamental framework, $\{g,\Gamma\}$\\
AF & Alternative framework, $\{g,\mathring{O}^N, A^{I}\}$\\
AF$^{\circ}$ & Diminished alternative framework, $\{g,\mathring{O}^N, B^{A}\}$\\
d-AF$^{\circ}$ & -, $\{g,\mathring{O}^I, B^{\mathcal{A}}\}$\\
\hline
\end{tabular}
\caption{\label{tab:Acronyms} Acronyms used in this work and their full name.}
\end{table}

\begin{table}[H]
\centering
\begin{tabular}{|l|l|}
\hline
Symbol&Definition\\
\hline 
$\mathring{O}_{\lambda\mu\nu}$ & $\{H_{\lambda\mu\nu},\mathring{t}_{\lambda\mu\nu}, \mathring{\pi}_{\lambda\mu\nu},\mathring{q}_{\lambda\mu\nu}\} $ \\
$A_\mu$ & $\{ T_\mu,\rho_\mu,u_\mu\}$ \\ 
$B_\mu$ in the AF$^{\circ}$ & \eqref{def:DAFBdoublet}\\
$B_\mu$ in the d-AF$^{\circ}$ & \eqref{def:DAFBtriplet}\\
$\mathsf{g}$ & $\det(g_{\mu\nu})$\\
$\mathring{\mathcal{O}}^{\lambda\mu\nu}_N$ & $\sqrt{-\mathsf{g}}^{-1}\delta I/\delta \mathring{O}^N_{\lambda\mu\nu}$\\
${\mathcal{A}}^{\mu}_I$ & $\sqrt{-\mathsf{g}}^{-1}\delta I/\delta A^I_{\mu}$\\
${\mathcal{B}}^{\mu}_{A/\mathcal{A}}$ & $\sqrt{-\mathsf{g}}^{-1}\delta I/\delta B^{A/\mathcal{A}}_{\mu}$\\
$\Delta_\lambda{}^{\mu\nu}$ &  $\sqrt{-\mathsf{g}}^{-1}\delta I/\delta \Gamma^\lambda{}_{\mu\nu}$\\
$E_{\mu\nu}$ & $\sqrt{-\mathsf{g}}^{-1}\delta I/\delta g_{\mu\nu}$ in the FF\\
$\hat{E}_{\mu\nu}$ & $\sqrt{-\mathsf{g}}^{-1}\delta I/\delta g_{\mu\nu}$ in the AF\\
$\check{E}_{\mu\nu}$ & $\sqrt{-\mathsf{g}}^{-1}\delta I/\delta g_{\mu\nu}$ in the d-AF$^{\circ}$\\
$R_T,R_V$ & \eqref{eq:AFscalConstituents}\\
$\mathcal{R}_V$ & \eqref{eq:DAFRV}\\
$\hat{R}_T, \hat{\mathcal{R}}_V$ & \eqref{eq:DAFscalConstituents}\\
\hline
\end{tabular}
\caption{\label{tab:Symbols} Some symbols used in this work and their definition.}
\end{table}
 
\acknowledgments

D.I. work is funded by the Estonian Research Council grant (SJD14). K. P. acknowledges financial support provided by the European Regional Development Fund (ERDF) through the Center of Excellence TK133 “The Dark Side of the Universe” and PRG356 “Gauge gravity: unification, extensions and phenomenology”. K.P. also acknowledges participation in the COST Association Action CA18108 “Quantum Gravity Phenomenology in the Multimessenger Approach (QG-MM)”. The authors would also like to thank Anastasios C. Petkou and Roberto Percacci for the fruitful discussions and valuable comments during this work.

\bibliographystyle{unsrt}
\bibliography{refs.bib}

\begin{thebibliography}{10}

\bibitem{Martin:2012bt}
Jerome Martin.
\newblock {Everything You Always Wanted To Know About The Cosmological Constant
  Problem (But Were Afraid To Ask)}.
\newblock {\em Comptes Rendus Physique}, 13:566--665, 2012.

\bibitem{PhysRevLett.22.1071}
Charles~W. Misner.
\newblock Mixmaster universe.
\newblock {\em Phys. Rev. Lett.}, 22:1071--1074, May 1969.

\bibitem{dicke1970gravitation}
R.H. Dicke.
\newblock {\em Gravitation and the Universe}.
\newblock American Philosophical Society: Memoirs of the American Philosophical
  Society. American Philosophical Society, 1970.

\bibitem{1979grec.conf..504D}
R.~H. {Dicke} and P.~J.~E. {Peebles}.
\newblock {The big bang cosmology - enigmas and nostrums.}
\newblock In S.~W. {Hawking} and W.~{Israel}, editors, {\em General Relativity:
  An Einstein centenary survey}, pages 504--517, January 1979.

\bibitem{Weinberg:1972kfs}
Steven Weinberg.
\newblock {\em {Gravitation and Cosmology}: {Principles and Applications of the
  General Theory of Relativity}}.
\newblock John Wiley and Sons, New York, 1972.

\bibitem{PhysRevD.23.347}
Alan~H. Guth.
\newblock Inflationary universe: A possible solution to the horizon and
  flatness problems.
\newblock {\em Phys. Rev. D}, 23:347--356, Jan 1981.

\bibitem{CANTATA:2021ktz}
Emmanuel~N. Saridakis et~al.
\newblock {Modified Gravity and Cosmology: An Update by the CANTATA Network}.
\newblock 5 2021.

\bibitem{clifton2012modified}
Timothy Clifton, Pedro~G Ferreira, Antonio Padilla, and Constantinos Skordis.
\newblock Modified gravity and cosmology.
\newblock {\em Physics reports}, 513(1-3):1--189, 2012.

\bibitem{Heisenberg:2018vsk}
Lavinia Heisenberg.
\newblock {A systematic approach to generalisations of General Relativity and
  their cosmological implications}.
\newblock {\em Phys. Rept.}, 796:1--113, 2019.

\bibitem{Lovelock:1971yv}
D.~Lovelock.
\newblock {The Einstein tensor and its generalizations}.
\newblock {\em J. Math. Phys.}, 12:498--501, 1971.

\bibitem{1972JMP....13..874L}
David {Lovelock}.
\newblock {The Four-Dimensionality of Space and the Einstein Tensor}.
\newblock {\em Journal of Mathematical Physics}, 13(6):874--876, June 1972.

\bibitem{HEHL19951}
Friedrich~W. Hehl, J.Dermott McCrea, Eckehard~W. Mielke, and Yuval Ne'eman.
\newblock Metric-affine gauge theory of gravity: field equations, noether
  identities, world spinors, and breaking of dilation invariance.
\newblock {\em Physics Reports}, 258(1):1--171, 1995.

\bibitem{HehlKerlickHeyde+1976+111+114}
Friedrich~W. Hehl, G.~David Kerlick, and Paul von~der Heyde.
\newblock On hypermomentum in general relativity i. the notion of
  hypermomentum.
\newblock {\em Zeitschrift für Naturforschung A}, 31(2):111--114, 1976.

\bibitem{HehlKerlickHeyde+1976+524+527}
Friedrich~W. Hehl, G.~David Kerlick, and Paul von~der Heyde.
\newblock On hypermomentum in general relativity ii. the geometry of spacetime.
\newblock {\em Zeitschrift für Naturforschung A}, 31(6):524--527, 1976.

\bibitem{HehlKerlickHeyde+1976+823+827}
Friedrich~W. Hehl, G.~David Kerlick, and Paul von~der Heyde.
\newblock On hypermomentum in general relativity iii. coupling hypermomentum to
  geometry.
\newblock {\em Zeitschrift für Naturforschung A}, 31(7):823--827, 1976.

\bibitem{HEHL1976446}
F.W. Hehl, G.D. Kerlick, and P.~{Von der Heyde}.
\newblock On a new metric affine theory of gravitation.
\newblock {\em Physics Letters B}, 63(4):446--448, 1976.

\bibitem{Percacci:2020ddy}
R.~Percacci and E.~Sezgin.
\newblock {New class of ghost- and tachyon-free metric affine gravities}.
\newblock {\em Phys. Rev. D}, 101(8):084040, 2020.

\bibitem{Percacci:2020bzf}
Roberto Percacci.
\newblock {Towards Metric-Affine Quantum Gravity}.
\newblock {\em Int. J. Geom. Meth. Mod. Phys.}, 17(supp01):2040003, 2020.

\bibitem{Pagani:2015ema}
Carlo Pagani and Roberto Percacci.
\newblock {Quantum gravity with torsion and non-metricity}.
\newblock {\em Class. Quant. Grav.}, 32(19):195019, 2015.

\bibitem{Iosifidis:2019dua}
Damianos Iosifidis.
\newblock {\em {Metric-Affine Gravity and Cosmology/Aspects of Torsion and
  non-Metricity in Gravity Theories}}.
\newblock PhD thesis, 2019.

\bibitem{Iosifidis:2020gth}
Damianos Iosifidis.
\newblock {Cosmological Hyperfluids, Torsion and Non-metricity}.
\newblock {\em Eur. Phys. J. C}, 80(11):1042, 2020.

\bibitem{Iosifidis:2021nra}
Damianos Iosifidis.
\newblock {The Perfect Hyperfluid of Metric-Affine Gravity: The Foundation}.
\newblock {\em JCAP}, 04:072, 2021.

\bibitem{Shimada:2018lnm}
Keigo Shimada, Katsuki Aoki, and Kei-ichi Maeda.
\newblock {Metric-affine Gravity and Inflation}.
\newblock {\em Phys. Rev. D}, 99(10):104020, 2019.

\bibitem{Mikura:2020qhc}
Yusuke Mikura, Yuichiro Tada, and Shuichiro Yokoyama.
\newblock {Conformal inflation in the metric-affine geometry}.
\newblock {\em EPL}, 132(3):39001, 2020.

\bibitem{BeltranJimenez:2020sqf}
Jose Beltr\'an~Jim\'enez and Adri\`a Delhom.
\newblock {Instabilities in metric-affine theories of gravity with higher order
  curvature terms}.
\newblock {\em Eur. Phys. J. C}, 80(6):585, 2020.

\bibitem{PhysRevD.101.044011}
Dietmar~Silke Klemm and Lucrezia Ravera.
\newblock Einstein manifolds with torsion and nonmetricity.
\newblock {\em Phys. Rev. D}, 101:044011, Feb 2020.

\bibitem{Iosifidis:2018zwo}
Damianos Iosifidis and Tomi Koivisto.
\newblock {Scale transformations in metric-affine geometry}.
\newblock {\em Universe}, 5:82, 2019.

\bibitem{Iosifidis:2021fnq}
Damianos Iosifidis and Lucrezia Ravera.
\newblock {Cosmology of quadratic metric-affine gravity}.
\newblock {\em Phys. Rev. D}, 105(2):024007, 2022.

\bibitem{Iosifidis:2021bad}
Damianos Iosifidis.
\newblock {The full quadratic metric-affine gravity (including parity odd
  terms): exact solutions for the affine-connection}.
\newblock {\em Class. Quant. Grav.}, 39(9):095002, 2022.

\bibitem{Rigouzzo:2022yan}
Claire Rigouzzo and Sebastian Zell.
\newblock {Coupling metric-affine gravity to a Higgs-like scalar field}.
\newblock {\em Phys. Rev. D}, 106(2):024015, 2022.

\bibitem{Jimenez-Cano:2022sds}
Alejandro Jim\'enez-Cano and Francisco~Jos\'e Maldonado~Torralba.
\newblock {Vector stability in quadratic metric-affine theories}.
\newblock {\em JCAP}, 09:044, 2022.

\bibitem{Bahamonde:2021akc}
Sebastian Bahamonde and Jorge Gigante~Valcarcel.
\newblock {Observational constraints in metric-affine gravity}.
\newblock {\em Eur. Phys. J. C}, 81(6):495, 2021.

\bibitem{BeltranJimenez:2020sih}
Jose Beltr\'an~Jim\'enez, Lavinia Heisenberg, and Tomi Koivisto.
\newblock {The coupling of matter and spacetime geometry}.
\newblock {\em Class. Quant. Grav.}, 37(19):195013, 2020.

\bibitem{Kubota:2020ehu}
Mio Kubota, Kin-Ya Oda, Keigo Shimada, and Masahide Yamaguchi.
\newblock {Cosmological Perturbations in Palatini Formalism}.
\newblock {\em JCAP}, 03:006, 2021.

\bibitem{Bahamonde:2022kwg}
Sebastian Bahamonde, Johann Chevrier, and Jorge Gigante~Valcarcel.
\newblock {New black hole solutions with a dynamical traceless nonmetricity
  tensor in Metric-Affine Gravity}.
\newblock 10 2022.

\bibitem{Bahamonde:2020fnq}
Sebastian Bahamonde and Jorge~Gigante Valcarcel.
\newblock {New models with independent dynamical torsion and nonmetricity
  fields}.
\newblock {\em JCAP}, 09:057, 2020.

\bibitem{Tresguerres:1995js}
R.~Tresguerres.
\newblock {Exact vacuum solutions of four-dimensional metric affine gauge
  theories of gravitation}.
\newblock {\em Z. Phys. C}, 65:347--354, 1995.

\bibitem{Tucker:1995fw}
Robin~W. Tucker and Charles Wang.
\newblock {Black holes with Weyl charge and nonRiemannian waves}.
\newblock {\em Class. Quant. Grav.}, 12:2587--2605, 1995.

\bibitem{Obukhov:1996pf}
Yu.~N. Obukhov, E.~J. Vlachynsky, W.~Esser, R.~Tresguerres, and F.~W. Hehl.
\newblock {An Exact solution of the metric affine gauge theory with dilation,
  shear, and spin charges}.
\newblock {\em Phys. Lett. A}, 220:1, 1996.

\bibitem{Cisterna:2020rkc}
Adolfo Cisterna, Gaston Giribet, Julio Oliva, and Konstantinos Pallikaris.
\newblock {Quasitopological electromagnetism and black holes}.
\newblock {\em Phys. Rev. D}, 101(12):124041, 2020.

\bibitem{abramowitz+stegun}
Milton Abramowitz and Irene~A. Stegun.
\newblock {\em Handbook of Mathematical Functions with Formulas, Graphs, and
  Mathematical Tables}.
\newblock Dover, New York, ninth dover printing, tenth gpo printing edition,
  1964.

\bibitem{Liu:2019rib}
Hai-Shan Liu, Zhan-Feng Mai, Yue-Zhou Li, and H.~L\"u.
\newblock {Quasi-topological Electromagnetism: Dark Energy, Dyonic Black Holes,
  Stable Photon Spheres and Hidden Electromagnetic Duality}.
\newblock {\em Sci. China Phys. Mech. Astron.}, 63:240411, 2020.

\bibitem{1966JETP...22..241S}
A.~D. {Sakharov}.
\newblock {The Initial Stage of an Expanding Universe and the Appearance of a
  Nonuniform Distribution of Matter}.
\newblock {\em Soviet Journal of Experimental and Theoretical Physics}, 22:241,
  January 1966.

\bibitem{Duff:1980qv}
M.~J. Duff and P.~van Nieuwenhuizen.
\newblock {Quantum Inequivalence of Different Field Representations}.
\newblock {\em Phys. Lett. B}, 94:179--182, 1980.

\end{thebibliography}


\end{document}